\documentclass[a4paper,12pt]{article}

\usepackage[english]{babel}
\usepackage[utf8x]{inputenc}
\usepackage[T1]{fontenc}
\usepackage[flushmargin]{footmisc}
\usepackage{setspace}
\usepackage[comma]{natbib}
\usepackage{float}
\usepackage{amsmath}
\usepackage{amsfonts}
\usepackage{amssymb}
\usepackage{caption}
\usepackage[a4paper,top=3cm,bottom=2cm,left=3cm,right=3cm,marginparwidth=1.75cm]{geometry}

\usepackage{graphicx}
\usepackage[colorlinks=true, allcolors=blue]{hyperref}
\def\ci{\perp\!\!\!\perp}

\begin{document} \doublespacing \pagestyle{plain}

\renewcommand*{\thefootnote}{\fnsymbol{footnote}}

\begin{center}
{\Large Direct and Indirect Effects based on Changes-in-Changes\footnote{We have benefited from comments by  Giuseppe Germinario as well as conference/seminar participants at the Universities of Neuch\^{a}tel, Melbourne, Sydney, Hamburg, and Lisbon, the Luxembourg Institute of Socio-Economic Research, the 2019 meeting of the Austro-Swiss Region of the International Biometric Society in Lausanne, and the 2019 meeting of the International Association for Applied Econometrics in Nicosia. Addresses for correspondence: Martin Huber, Chair of Applied Econometrics - Evaluation of Public Policies, University of Fribourg, Bd.\ de P\'{e}rolles 90, 1700 Fribourg, Switzerland, \href{mailto:Martin.Huber@unifr.ch}{Martin.Huber@unifr.ch}. Mark Schelker, Chair of Public Economics, University of Fribourg, Bd.\ de P\'{e}rolles 90, 1700 Fribourg, Switzerland, \href{mailto:Mark.Schelker@unifr.ch}{Mark.Schelker@unifr.ch}. Anthony Strittmatter, Swiss Institute for Empirical Economic Research (SEW), University of St.Gallen, Varnbüelstr. 14, 9000 St.Gallen, Switzerland, \href{mailto:Anthony.Strittmatter@unisg.ch}{Anthony.Strittmatter@unisg.ch}, \url{www.anthonystrittmatter.com}.}}

{\large \vspace{1.2 cm}}

{\large Martin Huber$^\dagger$, Mark Schelker$^\dagger$, Anthony Strittmatter$^\ddagger$}\smallskip

{\small {$^\dagger$University of Fribourg, Dept.\ of Economics}\\
\small {$^\ddagger$University of St.\ Gallen, Swiss Institute for Empirical Economic Research}\\
 \smallskip {\ \\[0pt]
\textbf{}}}\smallskip
\end{center}

\noindent \textbf{Abstract:} {\small We propose a novel approach for causal mediation analysis based on changes-in-changes assumptions restricting unobserved heterogeneity over time. This allows disentangling the causal effect of a binary treatment on a continuous outcome into an indirect effect operating through a binary intermediate variable (called mediator) and a direct effect running via other causal mechanisms. We identify average and quantile direct and indirect effects for various subgroups under the condition that the outcome is monotonic in the unobserved heterogeneity and that the distribution of the latter does not change over time conditional on the treatment and the mediator. We also provide a simulation study and an empirical application to the Jobs II programme.}\bigskip



\thispagestyle{empty}\pagebreak {\small \renewcommand{\thefootnote}{%
\arabic{footnote}} \setcounter{footnote}{0} \pagebreak %
\setcounter{footnote}{0} \pagebreak \setcounter{page}{1} }

\clearpage \renewcommand*{\thefootnote}{\arabic{footnote}}
\section{Introduction}

Causal mediation analysis aims at disentangling a total treatment effect into an indirect effect operating through an intermediate variable -- commonly referred to as mediator -- as well as the direct effect. The latter includes any causal mechanisms not operating through the mediator of interest. Even when the treatment is random, direct and indirect effects are generally not identified by simply controlling for the mediator without accounting for its potential endogeneity, as this likely introduces selection bias, see \cite{RoGr92}.

This paper suggests a novel identification strategy for causal mediation analysis based on changes-in-changes (CiC) as suggested by \cite{at06} for evaluating (total) average and quantile treatment effects. We adapt the approach to the identification of the direct effect and the indirect effect running through a binary mediator. The outcome variable must be continuous and is assumed to be observed both prior to and after treatment and mediator assignment as it is the case in repeated cross sections or panel data. The key identifying assumptions imply that the continuous outcome is strictly monotonic in unobserved heterogeneity and that the distribution of unobserved  heterogeneity does not change over time conditional on the treatment and the mediator (the latter assumption is also known as stationarity). Given appropriate common support conditions, this permits identifying direct effects on subpopulations conditional on the treatment and the mediator states, even if both treatment and mediator assignment are endogenous. 

Augmenting the assumptions by random treatment assignment and weak monotonicity of the mediator in the treatment allows for causal mediation analysis in subpopulations defined upon whether and how the mediator reacts to the treatment. 
Specifically, we show the identification of direct effects among those whose mediator is always one \citep[always-takers in the denomination of ][]{an96} and never one (never-takers) irrespective of treatment assignment, respectively. Furthermore, we identify the total, direct, and indirect treatment effects on those whose mediator value complies with treatment assignment (compliers). For any set of assumptions, we discuss the identification of both average and quantile direct and indirect effects. We note that if appropriately weighted, the respective average effects among compliers, always-takers, and never-takers add up to the average direct and indirect effects in the population.

Identification in the earlier mediation literature typically relied on linear models for the mediator and outcome equations and often neglected endogeneity issues, see for instance \cite{Co57}, \cite{JuKe81}, and \cite{BaKe86}. More recent contributions use more general identification approaches based on the potential outcome framework and take endogeneity issues explicitly into consideration. Examples include \cite{RoGr92}, \cite{Pearl01}, \cite{Robins2003},  \cite{PeSiva06}, \cite{VanderWeele09}, \cite{ImKeYa10}, \cite{Hong10}, \cite{AlbertNelson2011}, \cite{ImYa2011},  \cite{TchetgenTchetgenShpitser2011},  \cite{VansteelandtBekaertLange2012}, and \cite{Huber2012}. The vast majority of the literature assumes that the covariates observed in the data are sufficiently rich to control for treatment and mediator endogeneity. Also in empirical economics, there has been an increase in the application of such selection on observables approaches, see for instance \cite{SimonsenSkipper2006}, \cite{FlFl09}, \cite{HeckmanPintoSavelyev2013}, \cite{Huber2015}, \cite{Keeleetal2015}, \cite{ContiHeckmanPinto2016}, \cite{HuberLechnerMellace2017}, \cite{BijwaardJones2018},  \cite{BellaniBia2018}, \cite{HuberLechnerStrittmatter2018}, and \cite{doe19}. Comparably few studies in economics develop or apply instrumental variable approaches for disentangling direct and indirect effects, see for instance \cite{FrHu17}, \cite{PowdthaveeLekfuangfuWooden2013}, \cite{BrunelloFortSchneeweisWinterEbmer2016} and \cite{Chenetal2017}. Our paper provides another, CiC-based identification strategy that neither rests on selection on observables assumptions nor on instrumental variables  for the treatment or the mediator.

While most studies aim at evaluating direct and indirect effects in the total population, a smaller strand of the literature uses the principal stratification framework of \cite{FrangakisRubin02} to investigate effects in subpopulations (or principal strata) defined upon whether and how the mediator reacts to the treatment, see \cite{Ru04}. This approach has been criticized for typically focussing on direct effects on populations whose mediator is constant (i.e.\ always- and never-takers) rather than decomposing direct and indirect effects on compliers and for considering subpopulations rather than the total population, see \cite{VanderWeele08} and \cite{VanderWeele12}. \cite{Deuchertetal2017} suggest a difference-in-differences (DiD) strategy that alleviates such criticisms. Identification relies on a randomized treatment, monotonicity of the (binary) mediator in the treatment, and particular common trend assumptions on mean potential outcomes across principal strata. The latter imply that mean potential outcomes under specific treatment and mediator states change by the same amount over time across specific subpopulations. Depending on the strength of common trend and effect homogeneity assumptions across principal strata, direct and indirect effects are identified for different subpopulations and under the strongest set of assumptions even for the total population.

Our paper contributes to this literature on principal strata effects, but relies on different identifying assumptions than \cite{Deuchertetal2017}. While differential time trends across subpopulations are permitted, our approach restricts the conditional distribution of unobserved heterogeneity over time. The two sets of assumptions are not nested and their appropriateness is to be judged in the empirical context at hand. However, both approaches could be used simultaneously for testing the joint validity of the identifying assumptions of either method, in which case both CiC and DiD converge to the same, true average direct and indirect effects. As a further distinction to \cite{Deuchertetal2017}, our method also permits assessing quantile treatment effects (QTEs) rather than average effects only.

In independent work, \cite{saw19} proposes a CiC strategy to tackle non-compliance in randomized experiments when the exclusion restriction of random assignment is violated. While there is an overlap in some identification results of his study and ours (e.g.\ concerning the direct effect on never-takers), there are also important differences. First, \cite{saw19} predominantly focusses on the average treatment effect on the treated under one-sided non-compliance (ruling out always-takers), which then corresponds to the total effect on compliers. Our paper in addition disentangles the total complier effect into direct and indirect components. Second, under two-sided non-compliance (i.e.\ the existence of both never- and always-takers), \cite{saw19} identifies the total complier effect by assuming homogeneity of the direct effect, while we extend the CiC assumptions to the always-takers for identifying  (direct, indirect, and total) complier effects as well as the direct effect among always-takers. Third and in contrast to \cite{saw19}, we also provide identification results in the absence of randomization and monotonicity of the mediator in the treatment. On the other hand, \cite{saw19} in contrast to our study demonstrates that the CiC strategy  does not necessarily require pre-treatment outcomes, but may exploit any pre-treatment variable that has similar rank orders (as a function of unobserved heterogeneity) like the outcome of interest. 

We provide a simulation study in which we compare the CiC to the DiD approach to  illustrate our identification results. We also consider an empirical application to the Jobs II programme previously analysed by \cite{VinokurPriceSchul1995}, a randomized job training intervention designed to analyse the impact of job training on labour market and mental health outcomes. We investigate the direct effect of the randomized offer of treatment on a depression index, as well as its indirect effect through actual participation in the programme as mediator. The reason for investigating the direct effect is that treatment assignment could have a motivation or discouragement effect on those randomly offered or not offered the training. We, however, find the direct effect estimates to be close to zero and statistically insignificant and therefore no indication for the violation of the exclusion restriction when using treatment assignment as instrumental variable for actual participation. In contrast, the moderately negative total and indirect effects on those induced to participate by assignment are statistically significant at least at the 10\% level in all but one case and very much in line with the estimate obtained by instrumental variable regression.

The remainder of this study is organized as follows. Section \ref{effects} introduces the notation and defines the direct and indirect effects of interest. Section \ref{ident} presents the assumptions underlying our CiC approach as well as the identification results. Section \ref{sim} provides a simulation study. Section \ref{app} provides an application to Jobs II. Section \ref{con} concludes.

\section{Notation and effects}\label{effects}
\subsection{Average effects}

Let $D$ denote a binary treatment (e.g., receiving the offer to participate in a training programme) and $M$ a binary intermediate variable or mediator that may be a function of $D$ (e.g., the actual participation in a training programme). Furthermore, let $T$ indicate a particular time period: $T=0$ denotes the baseline period prior to the realisation of $D$ and $M$, $T=1$ the follow up period after measuring $D$ and $M$ in which the effect of the outcome is evaluated. Finally, let $Y_t$ denote the outcome of interest (e.g., health measures) in period $T=t$. Indexing the outcome by the time period $t \in \{0,1\}$ implies that it is measured both in the baseline period and after the realisation of $D$ and $M$. To define the parameters of interest, we make use of the potential outcome notation, see for instance \cite{rub74}, and denote by $Y_t(d,m)$ the potential outcome for treatment state $D=d$ and mediator state $M=m$ in time $T=t$, with $d, m, t, \in \{0,1\}$. Furthermore, let $M(d)$ denote the potential mediator as a function of the treatment state $d \in \{0,1\}$. For notational ease, we will not use any time index for $D$ and $M$, because either is assumed to be measured at a single point in time between $T=0$ and $T=1$, albeit not necessarily at the same point, as $D$ causally precedes $M$. Therefore, $D$ and $M$ correspond to the actual treatment and mediator status in $T=1$, while it is assumed that no treatment or mediation takes place in $T=0$.

Using this notation, the average treatment effect (ATE) in the ex-post period is defined as $\Delta_1 =E[Y_1(1,M(1))-Y_1(0,M(0))]$. That is, the ATE corresponds to the effect of $D$ on the outcome that either affects the latter directly (net of any effect on the mediator) or indirectly through an effect on $M$. Indeed, the total ATE can be disentangled into the direct and indirect effects, denoted by $\theta_1(d) = E[Y_1(1,M(d))-Y_1(0,M(d))]$ and $\delta_1(d)= E[Y_1(d,M(1)) -Y_1(d,M(0))]$, by adding and subtracting $Y_1(1,M(0))$ or $Y_1(0,M(1))$, respectively:
\begin{align*}
\Delta_1 & = E[Y_1(1,M(1))-Y_1(0,M(0))],\\
 & = \underbrace{E[Y_1(1,M(1))-Y_1(1,M(0))]}_{=\delta_1(1)}+\underbrace{E[Y_1(1,M(0))-Y_1(0,M(0))]}_{=\theta_1(0)},\\
 & = \underbrace{E[Y_1(1,M(1))-Y_1(0,M(1))]}_{ = \theta_1(1)}+\underbrace{E[Y_1(0,M(1))-Y_1(0,M(0))]}_{=\delta_1(0)}.
\end{align*}
Distinguishing between $\theta_1(1)$ and $\theta_1(0)$ or $\delta_1(1)$ and $\delta_1(0)$, respectively, implies the possibility of interaction effects between $D$ and $M$ such that the direct and indirect effects could be heterogeneous across values $d=1$ and $d=0$.

In our approach, we consider the concepts of direct and indirect effects within specific subpopulations. The latter are either defined conditional on the treatment and mediator values or conditional on potential mediator values under either treatment states, which matches the so-called principal stratum framework of \cite{FrangakisRubin02}. As outlined in \cite{an96} in the context of instrumental variable-based identification, any individual $i$ in the population belongs to one of four strata, henceforth denoted by $\tau$, according to their potential mediator status under either treatment state: always-takers ($a$: $M(1)=M(0)=1$) whose mediator is always one, compliers ($c$: $M(1)=1$, $M(0)=0$) whose mediator corresponds to the treatment value, defiers ($de$: $M(1)=0$, $M(0)=1$) whose mediator opposes the treatment value, and never-takers ($n$: $M(1)=M(0)=0$) whose mediator is never one. Note that $\tau$ cannot be pinned down for any individual, because either $M(1)$ or $M(0)$ is observed, but never both.


Let $\Delta_1^{\tau} = E[Y_1(1,M(1))-Y_1(0,M(0))|\tau]$ denote the ATE conditional on $\tau \in \{a,c,de,n\}$; $\theta_1^{\tau}(d)$ and $\delta_1^{\tau}(d)$ denote the corresponding direct and indirect effects. Because $M(1)=M(0)=0$ for any never-taker, the indirect effect for this group is by definition zero $(\delta_1^{n}(d)=E[Y_1(d,0) -Y_1(d,0)|\tau=n]=0)$ and $\Delta_1^{n} = E[Y_1(1,0)-Y_1(0,0)|\tau=n]=\theta_1^{n}(1)=\theta_1^{n}(0)=\theta_1^{n}$ equals the direct effect for never-takers. Correspondingly, because $M(1)=M(0)=1$ for any always-taker, the indirect effect for this group is by definition zero $(\delta_1^{a}(d)=E[Y_1(d,1) -Y_1(d,1)|\tau=a]=0)$ and $\Delta_1^{a} = E[Y_1(1,1)-Y_1(0,1)|\tau=a]=\theta_1^{a}(1)=\theta_1^{a}(0)=\theta_1^{a}$ equals the direct effect for always-takers. For the compliers, both direct and indirect effects may exist. Note that $M(d)=d$ due to the definition of compliers. Accordingly, $\theta_1^{c}(d) = E[Y_1(1,d)-Y_1(0,d)|\tau=c]$ equals the direct effect for compliers, $\delta_1^{c}(d)= E[Y_1(d,1) -Y_1(d,0)|\tau=c]$ equals the indirect effect for compliers, and $\Delta_1^{c}= E[Y_1(1,1) -Y_1(0,0)|\tau=c]$ equals the total effect for compliers. In the absence of any direct effect, the indirect effects on the compliers are homogeneous, $\delta_1^{c}(1)=\delta_1^{c}(0)=\delta_1^{c}$, and correspond to the local average treatment effect \citep[LATE, e.g.,][]{an96}. Analogous results hold for the defiers.

As already mentioned, we will also consider direct effects conditional on specific values $D=d$ and mediator states $M=M(d)=m$, which are denoted by $\theta_1^{d,m}(d)=E[Y_1(1,m)-Y_1(0,m)|D=d,M(d)=m]$. These parameters are identified under weaker assumptions than strata-specific effects, but are also less straightforward to interpret, as they refer to mixtures of two strata. For instance, $\theta_1^{1,0}(1)=E[Y_1(1,0)-Y_1(0,0)|D=1,M(1)=0]$ is the effect on a mixture of  never-takers and defiers, as these two groups satisfy $M(1)=0$. Likewise, $\theta_1^{0,0}(0)$ refers to never-takers and compliers satisfying $M(0)=0$,  $\theta_1^{0,1}(0)$ to always-takers and defiers satisfying $M(0)=1$, and $\theta_1^{1,1}(1)$ to always-takers and compliers satisfying $M(1)=1$.

\subsection{Quantile effects}

We denote by $F_{Y_{t}(d,m)}(y) = \Pr(Y_t(d,m) \leq y)$ the cumulative distribution function of $Y_t(d,m)$ at outcome level $y$. Its inverse, $F_{Y_{t}(d,m)}^{-1}(q) = \inf \{y : F_{Y_t(d,m)}(y) \geq q \}$, is the quantile function of $Y_t(d,m)$ at rank $q$. The total QTE are denoted by $\Delta_1(q) = F_{Y_1(1,M(1))}^{-1}(q) -F_{Y_1(0,M(0))}^{-1}(q)$.
The QTE can be disentangled into the direct quantile effects, denoted by $\theta_1(q,d) = F_{Y_1(1,M(d))}^{-1}(q) -F_{Y_1(0,M(d))}^{-1}(q)$, and the indirect quantile effects, denoted by $\delta_1(q,d) = F_{Y_1(d,M(1))}^{-1}(q) -F_{Y_1(d,M(0))}^{-1}(q)$.

The conditional distribution function in stratum $\tau$ is $F_{Y_{t}(d,m)|\tau}(y) = \Pr(Y_t(d,m) \leq y |\tau)$ and the corresponding conditional quantile function is $F_{Y_t(d,m)|\tau}^{-1}(q) = \inf \{y : F_{Y_{t}(d,m)|\tau}(y) \geq q \}$ for $\tau \in \{a,c,d,n\}$. Using the previously described stratification framework, we define the QTE conditional on $\tau \in \{a,c,de,n\}$: $\Delta_1^{\tau}(q) = F_{Y_1(1,M(1))|\tau}^{-1}(q)-F_{Y_1(0,M(0))|\tau}^{-1}(q)$. The direct quantile treatment effect among never-takers equals $\Delta_1^{n} (q)= F_{Y_1(1,0)|n}^{-1}(q)-F_{Y_1(0,0)|n}^{-1}(q) =\theta_1^{n}(q)$. The direct quantile effect among always-takers equals $\Delta_1^{a} (q)= F_{Y_1(1,1)|a}^{-1}(q)-F_{Y_1(0,1)|a}^{-1}(q) =\theta_1^{a}(q)$. The total QTE among compliers equals $\Delta_1^{c}(q) = F_{Y_1(1,1)|c}^{-1}(q)-F_{Y_1(0,0)|c}^{-1}(q)$, the direct quantile effect among compliers equals $\theta_1^{c}(q,d) = F_{Y_1(1,d)|c}^{-1}(q)-F_{Y_1(0,d)|c}^{-1}(q)$, and the indirect quantile effect among compliers equals $\delta_1^{c}(q,d) = F_{Y_1(d,1)|c}^{-1}(q)-F_{Y_1(d,0)|c}^{-1}(q)$. Finally, we define the direct quantile treatment effects conditional on specific values $D=d$ and mediator states $M=M(d)=m$,
\begin{align*}
\theta_1^{d,m}(q,1)&=F_{Y_1(1,m)|D=d,M(1)=m}^{-1}(q)-F_{Y_1(0,m)|D=d,M(1)=m}^{-1}(q) \mbox{ and} \\
\theta_1^{d,m}(q,0)&=F_{Y_1(1,m)|D=d,M(0)=m}^{-1}(q)-F_{Y_1(0,m)|D=d,M(0)=m}^{-1}(q),
\end{align*}
with the quantile function $F_{Y_t(d,m)|D=d,M(d)=m}^{-1}(q) = \inf \{y : F_{Y_{t}(d,m)|D=d,M(d)=m}(y) \geq q \}$ and the distribution function $F_{Y_{t}(d,m)|D=d,M(d)=m}(y) = \Pr(Y_t(d,m) \leq y |D=d,M(d)=m)$.

\subsection{Observed distribution and quantile transformations}

We subsequently define various functions of the observed data required for the identification results. The conditional distribution function of the observed outcome $Y_t$ conditional on treatment value $d$ and mediator state $m$, is given by $F_{Y_{t}|D=d,M=m}(y) = \Pr(Y_t \leq y |D=d,M=m)$ for $d,m \in \{0,1\}$. The corresponding conditional quantile function is $F_{Y_{t}|D=d,M=m}^{-1}(q) = \inf \{y : F_{Y_t|D=d,M=m}(y) \geq q \}$. Furthermore,
\begin{equation*}
Q_{dm}(y) := F_{Y_{1}|D=d,M=m}^{-1} \circ F_{Y_{0}|D=d,M=m}(y) = F_{Y_{1}|D=d,M=m}^{-1}(F_{Y_{0}|D=d,M=m}(y))
\end{equation*}
is the quantile-quantile transform of the conditional outcome from period 0 to 1 given treatment $d$ and mediator status $m$. This transform maps $y$ at rank $q$ in period 0 ($q = F_{Y_{0}|D=d,M=m}(y)$) into the corresponding $y'$ at rank $q$ in period 1 ($y'= F_{Y_{1}|D=d,M=m}^{-1}(q)$).

\section{Identification and Estimation}\label{ident}
\subsection{Identification}

This sections discusses the identifying assumptions along with the identification results for the various direct and indirect effects. We note that our assumptions could be adjusted to only hold conditional on a vector of observed covariates. In this case, the identification results would hold within cells defined upon covariate values. In our main discussion, however, covariates are not considered for the sake of ease of notation. For notational convenience, we maintain throughout that $\Pr(T=t, D=d, M=m)>0$ for $t,d,m$ $\in\{1,0\}$, implying that all possible treatment-mediator combinations exist in the population in both time periods. Our first assumption implies that potential outcomes are characterized by a continuous nonparametric function, denoted by $h$, that is strictly monotonic in a scalar $U$ that reflects unobserved heterogeneity.\vspace{5 pt}\\
\textbf{Assumption 1:} Strict monotonicity of continuous potential outcomes in unobserved heterogeneity.\\
The potential outcomes satisfy the following model: $Y_t(d,m)= h(d,m, t, U)$, with the general function $h$ being continuous and strictly increasing in the scalar unobservable $U \in \mathbb{R}$ for all $d,m,t \in \{0,1\}$.\vspace{5 pt}\\
Assumption 1 requires the potential outcomes to be continuous implying that there is a one-to-one correspondence between a potential outcome's distribution and quantile functions, which is a condition for point identification. For discrete potential outcomes, only bounds on the effects could be identified, in analogy to the discussion in \cite{at06} for total (rather than direct and indirect) effects. Assumption 1 also implies that individuals with identical unobserved characteristics $U$ have the same potential outcomes $Y_t(d,m)$, while higher values of $U$ correspond to strictly higher potential outcomes $Y_t(d,m)$. Strict monotonicity is automatically satisfied in additively separable models, but Assumption 1 also allows for more flexible non-additive structures that arise in nonparametric models.

The next assumption rules out anticipation effects of the treatment or the mediator on the outcome in the baseline period. This assumption is plausible if assignment to the treatment or the mediator cannot be foreseen in the baseline period, such that behavioral changes affecting the pre-treatment outcome are ruled out.\vspace{5 pt}\\
\textbf{Assumption 2:} No anticipation effect of $M$ and $D$ in the baseline period.\\
$Y_0(d,m) - Y_0(d',m') = 0\mbox{, for } d, d', m, m' \{1,0\}.$
\vspace{5 pt}\\
Similarly, \cite{at06} and \cite{cha15} assume the assignment to the treatment group does not affect the potential outcomes as long as the treatment is not yet realized. 

Furthermore, we assume conditional independence between unobserved heterogeneity and time periods given the treatment and no mediation. \vspace{5 pt}\\
\textbf{Assumption 3:} Conditional independence of $U$ and  $T$ given $D=1,M=0$ or $D=0,M=0$.\\
(a) $U\ci T|D=1,M=0$,\\
(b) $U\ci T|D=0,M=0$.\vspace{5 pt}\\
Under Assumption 3a, the distribution of $U$ is allowed to vary across groups defined upon treatment and mediator state, but not over time within the group with $D=1,M=0$. Assumption 3b imposes the same restriction conditional on $D=0,M=0$. Assumption 3 thus imposes stationarity of $U$ within groups defined on $D$ and $M$. This assumption is weaker than (and thus implied by) requiring that $U$ is constant across $T$ for each individual $i$.  For example, Assumption 3 is satisfied in the fixed effect model $U = \eta + v_t$, with $\eta$ being a time-invariant individual-specific unobservable (fixed effect) and $v_t$ an idiosyncratic time-varying unobservable with the same distribution in both time periods.

\cite{at06} and \cite{cha15} impose time invariance conditional on the treatment status, $U\ci T|D=d$, to identify the average treatment effect on the treated, $\varphi_1=E[Y_1(1,M(1))-Y_1(0,M(0))|D=1]$ or local average treatment effect, $\varphi_1=E[Y_1(1,M(1))-Y_1(0,M(0))|\tau=c]$, respectively. We additionally condition on the mediator status to identify direct and indirect effects. 

For our next assumption, we introduce some further notation. Let $F_{U|d,m}(u) )= \Pr(U \leq u |D=d,M=m)$ be the conditional distribution of $U$ with support $\mathbb{U}_{dm}$. \\ 
\textbf{Assumption 4:} Common support given $M=0$.\\
(a) $\mathbb{U}_{10}\subseteq \mathbb{U}_{00}$,\\
(b) $\mathbb{U}_{00}\subseteq \mathbb{U}_{10}$.\vspace{5 pt}\\
Assumption 4a is a common support assumption, implying that any possible value of $U$ in the population with $D=1,M=0$ is also contained in the population with $D=0,M=0$. Assumption 4b imposes that any value of $U$ conditional on $D=0,M=0$ also exists conditional on $D=1,M=0$. Both assumptions together imply that the support of $U$ is the same in both populations, albeit the distributions may generally differ.

Assumptions 1 to 3 permit identifying direct effects on mixed populations of never-takers and defiers as well as never-takers and compliers, respectively, as formally stated in Theorem 1.\\
\noindent \textbf{Theorem 1:} Under Assumptions 1–3,
\begin{itemize}
\item[(a)] and Assumption 4a, the average and quantile direct effects under $d=1$ conditional on $D=1$ and $M(1)=0$ are identified:
\begin{align*}
\theta_1^{1,0}(1)&= E[Y_1-Q_{00}(Y_0)|D=1,M=0], \\
\theta_1^{1,0}(q,1)&= F_{Y_1|D=1,M=0}^{-1}(q)-F_{Q_{00}(Y_0)|D=1,M=0}^{-1}(q).
\end{align*}
\item[(b)] and Assumption 4b, the average and quantile direct effects under $d=0$ conditional on $D=0$ and $M(0)=0$ are identified:
\begin{align*}
\theta_1^{0,0}(0)&= E[Q_{10}(Y_0)-Y_1|D=0,M=0], \\
\theta_1^{0,0}(q,0)&= F_{Q_{10}(Y_0)|D=0,M=0}^{-1}(q)-F_{Y_1|D=0,M=0}^{-1}(q).
\end{align*}
\end{itemize}
\textbf{Proof.} See Appendix \ref{appA}.

To identify direct effects on further populations, we invoke a conditional independence assumption that is in the spirit of Assumption 3, but refers to different combinations of the treatment and the mediator.\vspace{5 pt}\\
\textbf{Assumption 5:}  Conditional independence of $U$ and  $T$ given $D=0,M=1$ or $D=1,M=1$.\\
(a) $U\ci T|D=0,M=1$,\\
(b) $U\ci T|D=1,M=1$.\vspace{5 pt}\\
Under Assumption 5a, the distribution of $U$ is allowed to vary by treatment and mediator group, but not over time conditional on $D=0,M=1$. Assumption 5b imposes the same restriction conditional on $D=1,M=1$.

Assumption 6 is similar to Assumption 4, but imposes common support conditional on $M=1$ rather than  $M=0$.\vspace{5 pt}\\
\textbf{Assumption 6:} Common support given $M=1$.\\
(a) $\mathbb{U}_{01}\subseteq \mathbb{U}_{11}$,\\
(b) $\mathbb{U}_{11}\subseteq \mathbb{U}_{01}$.\vspace{5 pt}\\
Assumptions 6a implies that any possible value of $U$ in the population with $D=0,M=1$ is also contained in the population with $D=1,M=1$. Assumptions 6b states that any value of $U$ conditional on $D=1,M=1$ exists conditional on $D=0,M=1$.

Theorem 2 shows the identification of the direct effects on mixed populations of always-takers and defiers as well as always-takers and compliers.\\
\noindent \textbf{Theorem 2:} Under Assumptions 1-2, 5,
\begin{itemize}
\item[(a)] and Assumption 6a, the average and quantile direct effects under $d=1$ conditional on $D=0$ and $M(0)=1$ are identified:
\begin{align*}
\theta_1^{0,1}(0)&= E[Q_{11}(Y_0)-Y_1|D=0,M=1],\\
\theta_1^{0,1}(q,0)&= F_{Q_{11}(Y_0)|D=0,M=1}^{-1}(q) -F_{Y_1|D=0,M=1}^{-1}(q).
\end{align*}
\item[(b)] and Assumption 6b, the average and quantile direct effects under $d=1$ is identified conditional on $D=1$ and $M(1)=1$ are identified:
\begin{align*}
\theta_1^{1,1}(1)&= E[Y_1-Q_{01}(Y_0)|D=1,M=1],\\
\theta_1^{1,1}(q,1)&= F_{Y_1|D=1,M=1}^{-1}(q) - F_{Q_{01}(Y_0)|D=1,M=1}^{-1}(q).
\end{align*}
\end{itemize}
\textbf{Proof.} See Appendix \ref{appB}.

In the instrumental variable framework, any direct effects of the instrument are typically ruled out by imposing the exclusion restriction, in order to identify the causal effect of an endogenous regressor on the outcome, see for instance \cite{Imbens+94}. By considering $D$ as instrument and $M$ as endogenous regressor, $\theta_1^{1,0}(1)=\theta_1^{0,0}(0)=\theta_1^{0,1}(0)=\theta_1^{1,1}(1)=0$ yield testable implications of the exclusion restriction under Assumptions 1-6.


So far, we did not impose exogeneity of the treatment or mediator. In the following, we assume treatment exogeneity by invoking independence between the treatment and the potential post-treatment variables.\vspace{5 pt}\\
\textbf{Assumption 7:} Independence of the treatment and potential mediators/outcomes.\\
$ \{Y_t(d,m),M(d)\} \ci D \mbox{, for all } d,m,t, \in \{0,1\}.$\vspace{5 pt}\\
Assumption 7 implies that there are no confounders jointly affecting the treatment on the one hand and the mediator and/or outcome on the other hand. It is satisfied under treatment randomization as in successfully conducted experiments. This allows identifying the ATE: $\Delta_1 = E[Y_1|D=1] -E[Y_1|D=0]$.

Furthermore, we assume the mediator to be weakly monotonic in the treatment.\vspace{5 pt}\\
\textbf{Assumption 8:} Weak monotonicity of the mediator in the treatment.\\
$\Pr(M(1) \geq M(0)) =1.$\vspace{5 pt}\\
Assumption 8 is standard in the instrumental variable literature on local average treatment effects when denoting by $D$ the instrument and by $M$ the endogenous regressor, see \cite{Imbens+94} and \cite{an96}. It rules out the existence of defiers.

As discussed in the Appendix \ref{appC}, the total ATE $\Delta_1= E[Y_1 |D=1]- E[Y_1 |D=0]$ and QTE $\Delta_1(q) = F_{Y_{1} |D=1}^{-1}(q)- F_{Y_{1} |D=0}^{-1}(q)$ for the entire population are identified under Assumption 7. Furthermore, Assumptions 7 and 8 yield the strata proportions, denoted by $p_{\tau}= \Pr(\tau)$, as functions of the conditional mediator probabilities given the treatment, which we denote by $p_{(m|d)}=\Pr(M=m|D=d)$ for $d, m \in \{0,1\}$ (see Appendix \ref{appC}):
\begin{equation} \label{eq002a}
p_a =p_{1|0}, p_c = p_{1|1}-p_{1|0} = p_{0|0}-p_{0|1}, p_n = p_{0|1}.
\end{equation}
Furthermore, Assumptions 2, 7, and 8 imply that (see Appendix \ref{appC})
\begin{equation} \label{eq002c}
\Delta_{0,c} =E[Y_0(1,1) - Y_0(0,0)|c] = \frac{E[Y_0|D=1] -E[Y_0|D=0] }{p_{1|1} - p_{1|0}} =  0.
\end{equation}
Therefore, a rejection of the testable implication $E[Y_0|D=1] -E[Y_0|D=0]=0$ in the data would point to a violation of these assumptions.

Assumptions 7 and 8 permit identifying additional parameters, namely the total, direct, and indirect effects on compliers, and the direct effects on never- and always-takers, as shown in Theorems 3 to 5. This follows from the fact that defiers are ruled out and that the proportions and potential outcome distributions of the various principal strata are not selective w.r.t.\ the treatment.\\
\noindent \textbf{Theorem 3:} Under Assumptions 1–3, 7-8,
\begin{itemize}
\item[a)] and Assumption 4a, the average and quantile direct effects on never-takers are identified:
\begin{equation*}
\theta_1^n= \theta_1^{1,0}(1) \mbox{ and } \theta_1^n(q)= \theta_1^{1,0}(q,1).
\end{equation*}
\item[b)] and Assumption 4, the average direct effect under $d = 0$ on compliers is identified:
\begin{align*}
\displaystyle \theta_1^{c}(0) =& \frac{p_{0|0}}{p_{0|0} - p_{0|1}} \theta_1^{0,0}(0)   - \frac{p_{0|1}}{p_{0|0} - p_{0|1}} \theta_1^{1,0}(1) .
\end{align*}
Furthermore, the potential outcome distributions under $d = 0$ on compliers are identified:
\begin{equation} \begin{array}{rl} \displaystyle
 F_{Y_{1}(1,0)|\tau=c}(y) =& \displaystyle \frac{p_{0|0}}{p_{0|0} - p_{0|1}}  F_{Q_{10}(Y_{0})|D=0,M=0}(y) \\& \displaystyle \qquad - \frac{p_{0|1}}{ p_{0|0} - p_{0|1}c}  F_{Y_1|D=1,M=0}(y), \label{dd1}\end{array}
 \end{equation}
 \begin{equation}
 \begin{array}{rl}\displaystyle
F_{Y_{1}(0,0)|\tau=c}(y) =& \displaystyle \frac{p_{0|0}}{p_{0|0} - p_{0|1}} F_{Y_{1}|D=0,M=0}(y) \\& \displaystyle \qquad - \frac{p_{0|1} }{p_{0|0} - p_{0|1}}F_{Q_{00}(Y_{0})|D=1,M=0}(y)  . \label{dd2}
\end{array}
\end{equation}
Therefore, the direct quantile effect under $d = 0$ on compliers, $\theta_1^{c}(q,0) = F_{Y_{1}(1,0)|c}^{-1}(q)-F_{Y_{1}(0,0)|c}^{-1}(q)$, is identified.
\end{itemize}
\textbf{Proof.} See Appendix \ref{appD}.

\noindent \textbf{Theorem 4:} Under Assumptions 1–2, 5, 7-8,
\begin{itemize}
\item[a)] and Assumption 6a, the average and quantile direct effects on
always-takers are identified:
\begin{equation*}
\theta_1^a= \theta_1^{0,1}(0) \mbox{ and } \theta_1^a(q)= \theta_1^{0,1}(q,0).
\end{equation*}
\item[b)] and Assumption 6, the average direct effect under $d = 1$ on compliers is identified:
\begin{align*}
\theta_{1}^{c}(1) =& \frac{p_{1|1}}{ p_{1|1} - p_{1|0}} \theta_1^{1,1}(1) -\frac{p_{1|0}}{p_{1|1} - p_{1|0}}\theta_1^{0,1}(0).
\end{align*}
Furthermore, the potential outcome distributions under $d = 1$ for compliers are identified:
\begin{equation} \begin{array}{rl} \displaystyle
 F_{Y_{1}(1,1)|\tau=c}(y) =& \displaystyle \frac{p_{1|1}}{p_{1|1} - p_{1|0}} F_{Y_{1}|D=1,M=1}(y) \\& \displaystyle \qquad - \frac{p_{1|0} }{p_{1|1} - p_{1|0}}F_{Q_{11}(Y_0)|D=0,M=1}(y), \label{dd3} \end{array}
 \end{equation}
 \begin{equation}
 \begin{array}{rl}\displaystyle
F_{Y_{1}(0,1)|\tau=c}(y) =&\displaystyle \frac{p_{1|1}}{p_{1|1} - p_{1|0}}  F_{Q_{01}(Y_{0})|D=1,M=1}(y)\\ & \displaystyle \qquad - \frac{p_{1|0}}{ p_{1|1} - p_{1|0}}  F_{Y_1|D=0,M=1}(y). \label{dd4} \end{array}
\end{equation}
Therefore, the direct quantile effect under $d = 1$ on compliers $\theta_1^{c}(q,1) = F_{Y_{1}(1,1)|c}^{-1}(q)-F_{Y_{1}(0,1)|c}^{-1}(q)$ is identified.
\end{itemize}
\textbf{Proof.} See Appendix \ref{appE}.

\noindent \textbf{Theorem 5:} Under Assumptions 1-3, 5, 7-8,
\begin{itemize}
\item[a)] and Assumptions 4a, 6a, the total average treatment effect on compliers is identified:
\begin{align*}
\Delta_1^c=& \frac{p_{1|1}}{ p_{1|1} - p_{1|0}} E[Y_1|D=1,M=1] -\frac{p_{1|0}}{p_{1|1} - p_{1|0}}E[Q_{11}(Y_0)|D=0,M=1] \\
& − \frac{p_{0|0}}{ p_{1|1} - p_{1|0}} E[Y_1|D=0,M=0] +\frac{p_{0|1}}{p_{1|1} - p_{1|0}}E[Q_{00}(Y_0)|D=1,M=0].
\end{align*}
Furthermore, the total quantile treatment effect on compliers $\Delta_1^{c}(q) = F_{Y_{1}(1,1)|c}^{-1}(q)-F_{Y_{1}(0,0)|c}^{-1}(q)$ is identified using the inverse of (\ref{dd3}) and (\ref{dd2}).
\item[b)] and Assumptions 4a, 6b, the average indirect effect
under $d = 0$ on compliers is identified:
\begin{align*}
\delta_1^c(0) =& \frac{p_{1|1}}{p_{1|1}-p_{1|0}}E[Q_{01}(Y_0)|D=1,M=1] - \frac{p_{1|0}}{p_{1|1}-p_{1|0}}E[Y_1|D=0,M=1]\\
&\hspace{-0.15cm}- \frac{p_{0|0}}{p_{1|1} - p_{1|0}} E[Y_1|D=0,M=0] +\frac{p_{0|1}}{p_{1|1} - p_{1|0}}E[Q_{00}(Y_0)|D=1,M=0].
\end{align*}
Furthermore, the quantile indirect effect
under $d = 0$ on compliers $\delta_1^{c}(q,0) = F_{Y_{1}(0,1)|c}^{-1}(q)-F_{Y_{1}(0,0)|c}^{-1}(q)$ is identified using the inverse of (\ref{dd4}) and (\ref{dd2}).
\item[c)] and Assumptions 4b, 6a, the average indirect effect
under $d = 1$ on compliers is identified:
\begin{align*}
\delta_1^{c}(1) =& \frac{p_{1|1}}{ p_{1|1} - p_{1|0}} E[Y_1|D=1,M=1] -\frac{p_{1|0}}{p_{1|1} - p_{1|0}}E[Q_{11}(Y_0)|D=0,M=1]\\
&\hspace{-0.15cm}- \frac{p_{0|0}}{p_{1|1} - p_{1|0}}E[Q_{10}(Y_0)|D=0,M=0] + \frac{p_{0|1}}{p_{1|1} - p_{1|0}}E[Y_1|D=1,M=0].
\end{align*}
Furthermore, the quantile indirect effect
under $d = 1$ on compliers $\delta_1^{c}(q,1) = F_{Y_{1}(1,1)|c}^{-1}(q)-F_{Y_{1}(1,0)|c}^{-1}(q)$ is identified using the inverse of (\ref{dd3}) and (\ref{dd1}).
\end{itemize}
\textbf{Proof.} See Appendix \ref{appF}.

\subsection{Estimation}

 As in Assumption 5.1 of \cite{at06}, we assume standard regularity conditions, namely that conditional on $T=t$, $D=d$, and $M=m$, $Y$ is a random draw from that subpopulation defined in terms of $t,d,m$ $\in$ $\{1,0\}$. Furthermore, the outcome in the subpopulations required for the identification results of interest must have compact support and a density that is bounded from above and below as well as continuously differentiable. Denote by $N$ the total sample size across both periods and all treatment-mediator combinations and by $i$ $\in$ $\{1,...,N\}$ an index for the sampled subject, such that $(Y_i,D_i,M_i,T_i$) correspond to sample realizations of the random variables $(Y,D,M,T$).

 The total, direct, and indirect effects may be estimated using the sample analogy principle, which replaces population moments with sample moments \citep[e.g.][]{ma88}. For instance, any conditional mediator probability given the treatment, $\Pr(M=m|D=d)$, is to be replaced by an estimate thereof in the sample, $ \frac{\sum_{i=1}^{N} I\{M_i=m,D_i=d\}}{\sum_{i=1}^{N} I\{D_i=d\}}$. A crucial step is the estimation of the quantile-quantile transforms. The application of such quantile transformations dates at least back to \cite{juh91}, see also \cite{cha15}, \cite{wue19}, and \cite{str19} for recent applications. First, it requires estimating the conditional outcome distribution, $F_{Y_t|D=d,M=m}(y)$, by the conditional empirical distribution $\hat{F}_{Y_t|D=d,M=m}(y)=\frac{1}{\sum_{i=1}^{n}I\{D_i=d, M_i=m, T_i=t\}}\sum_{i:D_i=d,M_i=m,T_i=t}I\{Y_i\leq y\}$. Second, inverting the latter yields the empirical quantile function $\hat{F}_{Y_t|D=d,M=m}^{-1}(q)$. The empirical quantile-quantile transform is then obtained by
  \begin{eqnarray*}
 \hat{Q}_{dm}(y) = \hat{F}_{Y_{1}|D=d,M=m}^{-1}(\hat{F}_{Y_{0}|D=d,M=m}(y)).
  \end{eqnarray*}
 This permits estimating the average and quantile effects of interest. Average effects are estimated by replacing any (conditional) expectations with the corresponding sample averages in which the estimated quantile-quantile transforms enter as plug-in estimates. Taking $\theta_1^{1,0}$ (see Theorem 1) as an example, an estimate thereof is
 \begin{eqnarray*}
\hat{\theta}_1^{1,0}(1)&=&\frac{1}{\sum_{i=1}^{n}I\{D_i=1,M_i=0,T_i=1\}}\sum_{i:D_i=1,M_i=0,T_i=1}Y_i\\ &-&\frac{1}{\sum_{i=1}^{n}I\{D_i=1,M_i=0,T_i=0\}}\sum_{i:D_i=1,M_i=0,T_i=0}\hat{Q}_{00}(Y_i).
 \end{eqnarray*}
Likewhise, quantile effects are estimated based on the empirical quantiles.

For the estimation of total ATE and QTE, \cite{at06} show that the resulting estimators are $\sqrt{N}$-consistent and asymptotically normal, see their Theorems 5.1 and 5.3. These properties also apply to our context when splitting the sample into subgroups based on the values of a binary treatment and mediator (rather than the treatment only). For instance, the implications of Theorem 1 in \cite{at06} when considering subsamples with $D=1$ and $D=0$ carry over to considering subsamples with $D=1, M=0$ and $D=0, M=0$ for estimating the average direct effect on never-takers. In contrast to \cite{at06}, however, some of our identification results include the conditional mediator probabilities $\Pr(M=m|D=d)$. As the latter are estimated with $\sqrt{N}$-consistency, too, it follows that the resulting effect estimators are again $\sqrt{N}$-consistent and asymptotically normal. 
We use a non-parametric bootstrap approach to calculate the standard errors. \cite{cha15} show the validity of the bootstrap approach for such kind of estimators, which follows from their asymptotic normality.

For the case that identifying assumptions to only hold conditional on observed covariates, denoted by $X$, estimation must be adapted to allow for control variables. Following a suggestion by \cite{at06} in their Section 5.1, basing estimation on outcome residuals in which the association of $X$ and $Y$ has been purged by means of a regression is consistent under the additional assumption that the effects of $D$ and $M$ are homogeneous across covariates. As an alternative, \cite{mel15} propose a flexible semiparametric estimator that does not impose such a homogeneity-in-covariates assumption and show $\sqrt{N}$-consistency and asymptotic normality.

\section{Simulations}\label{sim}

To shape the intuition for our identification results, this section presents a brief simulation based on the following data generating process (DGP):
\begin{equation*}\label{sim}
T \sim Binom(0.5),\textrm{ }D \sim Binom(0.5),\textrm{ }U\sim Unif(-1,1),\textrm{ } V\sim N(0,1)
\end{equation*}
independent of each other, and
\begin{equation*}
M=I\{D+U+V>0\},\qquad Y_T=\Lambda((1+D+M+D\cdot M)\cdot T+U).
\end{equation*}
Treatment $D$ as well as the observed time period $T$ are randomized, while the mediator-outcome association is confounded due to the unobserved time constant heterogeneity $U$. The potential outcome in period $1$ is given by $Y_1(d,M(d'))=\Lambda((1+d+M(d')+d\cdot M(d'))+U)$, where $\Lambda$ denotes a link function. If the latter corresponds to the identity function,  our model is linear and implies a homogeneous time trend $T$ equal to 1. If $\Lambda$ is nonlinear, the time trend is heterogeneous, which invalidates the common trend assumption of difference-in-differences models. $M$ is not only a function of $D$ and $U$, but also of the unobserved random term $V$, which guarantees common support w.r.t.\ $U$, see Assumptions 4 and 6. Compliers, always-takers, and never-takers satisfy, respectively: $c=I\{U+V\leq 0, 1+U+V>0\}$, $a=I\{U+V>0\}$, and $n=I\{1+U+V\leq0\}$.

In the simulations with 1,000 replications, we consider two sample sizes ($N=1,000, 4,000$) and investigate the behaviour of our change-in-changes methods as well as the difference-in-differences approach of \cite{Deuchertetal2017} in both a linear ($\Lambda$ equal to identity function) and nonlinear outcome model where $\Lambda$ equals the exponential function. To implement the change-in-changes estimators in the simulations as well as the application in Section \ref{app}, we make use of the `cic' command in the \texttt{qte} R-package by \cite{Callaway2016} with its default values.

\begin{table}[!ht]
\caption{\label{simlinran} Linear model with random treatment}
\begin{center}
\begin{tabular}{lccccccc}
  \hline\hline
  & $\hat{\theta}_1^n$ & $\hat{\theta}_1^a$ & $\hat{\Delta}_c$ & $\hat{\theta}_1^c(1)$ & $\hat{\theta}_1^c(0)$ & $\hat{\delta}_1^c(1)$ & $\hat{\delta}_1^c(0)$  \\
  \hline
   \multicolumn{8}{c}{\textbf{A. Changes-in-Changes} }  \\\hline
   &\multicolumn{7}{c}{$N$=1,000} \\
  \hline
  bias & 0.00 & -0.00 & -0.01 & -0.01 & -0.01 & -0.00 & -0.01 \\
  sd & 0.11 & 0.08 & 0.23 & 0.10 & 0.13 & 0.27 & 0.27  \\
  rmse & 0.11 & 0.08 & 0.23 & 0.10 & 0.13 & 0.27 & 0.27  \\
  true & 1.00 & 2.00 & 3.00 & 2.00 & 1.00 & 2.00 & 1.00  \\
  relr & 0.11 & 0.04 & 0.08 & 0.05 & 0.13 & 0.14 & 0.27  \\
   \hline
  & \multicolumn{7}{c}{$N$=4,000} \\
  \hline
 bias & -0.00 & -0.00 & 0.00 & -0.00 & -0.01 & 0.01 & 0.01   \\
  sd & 0.06 & 0.04 & 0.12 & 0.05 & 0.07 & 0.14 & 0.14  \\
  rmse & 0.06 & 0.04 & 0.12 & 0.05 & 0.07 & 0.14 & 0.14  \\
  true & 1.00 & 2.00 & 3.00 & 2.00 & 1.00 & 2.00 & 1.00  \\
  relr & 0.06 & 0.02 & 0.04 & 0.02 & 0.07 & 0.07 & 0.14  \\
   \hline
        \multicolumn{8}{c}{\textbf{B. Difference-in-Differences} }  \\ \hline
      &\multicolumn{7}{c}{$N$=1,000} \\
  \hline
  bias &  0.01 & -0.00 & -0.01 & -0.01 & 0.00 & -0.02 & 0.00 \\
  sd &  0.11 & 0.09 & 0.14 & 0.14 & 0.12 & 0.19 & 0.10 \\
  rmse & 0.11 & 0.09 & 0.14 & 0.14 & 0.12 & 0.19 & 0.10 \\
  true &  1.00 & 2.00 & 3.00 & 2.00 & 1.00 & 2.00 & 1.00 \\
  relr &  0.11 & 0.04 & 0.05 & 0.07 & 0.12 & 0.10 & 0.10 \\
   \hline
  & \multicolumn{7}{c}{$N$=4,000} \\
  \hline
 bias  & -0.00 & -0.00 & 0.00 & -0.00 & -0.00 & 0.00 & 0.00 \\
  sd &  0.06 & 0.04 & 0.07 & 0.07 & 0.06 & 0.10 & 0.05 \\
  rmse & 0.06 & 0.04 & 0.07 & 0.07 & 0.06 & 0.10 & 0.05 \\
  true & 1.00 & 2.00 & 3.00 & 2.00 & 1.00 & 2.00 & 1.00 \\
  relr &   0.06 & 0.02 & 0.02 & 0.04 & 0.06 & 0.05 & 0.05 \\
   \hline
\end{tabular}
\end{center}
\par
\footnotesize{ Note: `bias', `sd', and `rmse' provide the bias, standard deviation, and root mean squared error of the respective estimator. `true' and `relr' are the respective true effect as well as the root mean squared error relative to the true effect.}
\end{table}

Table \ref{simlinran} reports the bias, standard deviation (`sd'), root mean squared error (`rmse'), true effect (`true'), and the relative root mean squared error in percent of the true effect (`relr')  of the respective estimators of $\theta_1^n$, $\theta_1^a$, $\Delta_c$,  $\theta_1^c(1)$, $\theta_1^c(0)$, $\delta_1^c(1)$, and $\delta_1^c(0)$ for the linear model. In this case, the identifying assumptions underlying both the change-in-changes (Panel A.) and difference-in-differences (Panel B.) estimators are satisfied. Specifically, the homogeneous time trend on the individual level satisfies any of the common trend assumptions in \cite{Deuchertetal2017}, while the monotonicity of $Y$ in $U$ and the independence of $T$ and $U$ satisfies the key assumptions of this paper. For this reason any of the estimates in Table \ref{simlinran} are close to being unbiased and appear to converge to the true effect at the parametric rate when comparing the results for the two different sample sizes.

\begin{table}[!ht]
\caption{\label{simnonlinran} Nonlinear model with random treatment}
\begin{center}
\begin{tabular}{lccccccc}
  \hline\hline
 & $\hat{\theta}_1^n$ & $\hat{\theta}_1^a$ & $\hat{\Delta}_c$ & $\hat{\theta}_1^c(1)$ & $\hat{\theta}_1^c(0)$ & $\hat{\delta}_1^c(1)$ & $\hat{\delta}_1^c(0)$  \\
  \hline
   \multicolumn{8}{c}{\textbf{A. Change-in-Changes} } \\ \hline
   &\multicolumn{7}{c}{$N$=1,000} \\
  \hline
  bias & 0.01 & -0.14 & -0.48 & -0.35 & -0.11 & -0.37 & -0.13  \\
  sd & 0.48 & 5.08 & 8.47 & 6.20 & 1.16 & 8.64 & 4.23  \\
  rmse & 0.48 & 5.08 & 8.48 & 6.21 & 1.17 & 8.65 & 4.23  \\
  true & 3.49 & 68.09 & 52.42 & 47.70 & 4.72 & 47.70 & 4.72  \\
  relr & 0.14 & 0.07 & 0.16 & 0.13 & 0.25 & 0.18 & 0.90  \\
  \hline
  & \multicolumn{7}{c}{$N$=4,000} \\
  \hline
bias & -0.01 & 0.01 & -0.00 & -0.11 & -0.07 & 0.07 & 0.11  \\
  sd & 0.25 & 2.63 & 4.37 & 3.20 & 0.66 & 4.44 & 2.04  \\
  rmse & 0.25 & 2.63 & 4.37 & 3.20 & 0.66 & 4.44 & 2.04  \\
  true & 3.49 & 68.09 & 52.45 & 47.73 & 4.72 & 47.73 & 4.72 \\
  relr & 0.07 & 0.04 & 0.08 & 0.07 & 0.14 & 0.09 & 0.43  \\
  \hline
\multicolumn{8}{c}{\textbf{B. Difference-in-Differences}} \\ \hline
   &\multicolumn{7}{c}{$N$=1,000} \\
  \hline
  bias &  -0.27 & -8.91 & 14.42 & 11.46 & -1.49 & 15.91 & 2.96 \\
  sd &  0.46 & 2.62 & 2.58 & 2.62 & 0.47 & 2.61 & 0.47 \\
  rmse &  0.53 & 9.29 & 14.65 & 11.76 & 1.56 & 16.12 & 2.99 \\
  true &  3.49 & 68.09 & 52.42 & 47.70 & 4.72 & 47.70 & 4.72 \\
  relr &  0.15 & 0.14 & 0.28 & 0.25 & 0.33 & 0.34 & 0.63 \\
  \hline
  & \multicolumn{7}{c}{$N$=4,000} \\
  \hline
bias & -0.28 & -8.79 & 14.51 & 11.57 & -1.51 & 16.02 & 2.94 \\
  sd &  0.24 & 1.28 & 1.26 & 1.28 & 0.25 & 1.27 & 0.23 \\
  rmse & 0.37 & 8.88 & 14.57 & 11.64 & 1.53 & 16.07 & 2.95 \\
  true &  3.49 & 68.09 & 52.45 & 47.73 & 4.72 & 47.73 & 4.72 \\
  relr &  0.11 & 0.13 & 0.28 & 0.24 & 0.32 & 0.34 & 0.62 \\
   \hline
\end{tabular}
\end{center}
\par
{\footnotesize Note: `bias', `sd', and `rmse' provide the bias, standard deviation, and root mean squared error of the respective estimator. `true' and `relr' are the respective true effect as well as the root mean squared error relative to the true effect.}
\end{table}

Table \ref{simnonlinran} provides the results for the exponential outcome model, in which the time trend is heterogeneous and interacts with $U$ through the nonlinear link function. While the change-in-changes assumptions hold (Panel A.), average time trends are heterogeneous across complier types such that the difference-in-differences approach (Panel B.) of \cite{Deuchertetal2017} is inconsistent. Accordingly, the biases of the change-in-changes estimates generally approach zero as the sample size increases, while this is not the case for the difference-in-differences estimates. Change-in-changes yields a lower root mean squared error than the respective difference-in-differences estimator in all but one case (namely $\hat{\delta}_1^c(0)$ with $N=1,000$) and its relative attractiveness increases in the sample size due to its lower bias.


\begin{table}[!ht]
\caption{\label{simnonlinnonran} Nonlinear model with non-random treatment}
\begin{center}
\begin{tabular}{lccccccc}
  \hline \hline
 & $\hat{\theta}_1^{0,1}$ & $\hat{\theta}_1^{1,0}$ & $\hat{\Delta}_c$ & $\hat{\theta}_1^c(1)$ & $\hat{\theta}_1^c(0)$ & $\hat{\delta}_1^c(1)$ & $\hat{\delta}_1^c(0)$  \\
  \hline
  \multicolumn{8}{c}{\textbf{A. Change-in-Changes} }\\ \hline
   &\multicolumn{7}{c}{$N$=1,000} \\
  \hline
  bias & 0.02 & 0.13 & 47.21 & 40.19 & -1.44 & 48.64 & 7.02 \\
  sd & 0.71 & 4.56 & 5.45 & 4.11 & 0.75 & 5.53 & 2.92 \\
  rmse & 0.71 & 4.56 & 47.52 & 40.40 & 1.62 & 48.96 & 7.60 \\
  true & 4.41 & 54.19 & 52.42 & 47.70 & 4.72 & 47.70 & 4.72  \\
  relr & 0.16 & 0.08 & 0.91 & 0.85 & 0.34 & 1.03 & 1.61  \\
  \hline
  & \multicolumn{7}{c}{$N$=4,000} \\
  \hline
  bias & -0.00 & 0.06 & 47.38 & 40.13 & -1.53 & 48.91 & 7.25  \\
  sd & 0.38 & 2.35 & 2.84 & 2.04 & 0.38 & 2.86 & 1.51  \\
  rmse & 0.38 & 2.35 & 47.47 & 40.18 & 1.57 & 48.99 & 7.40  \\
  true & 4.40 & 54.18 & 52.45 & 47.73 & 4.72 & 47.73 & 4.72 \\
  relr & 0.09 & 0.04 & 0.90 & 0.84 & 0.33 & 1.03 & 1.57  \\
    \hline
    \multicolumn{8}{c}{\textbf{B. Difference-in-Differences}} \\ \hline
   &\multicolumn{7}{c}{$N$=1,000} \\
  \hline
  bias &  0.35 & 19.98 & 29.00 & 27.65 & 0.04 & 28.96 & 1.35 \\
  sd & 0.67 & 2.48 & 2.46 & 2.48 & 0.67 & 2.51 & 0.45 \\
  rmse &  0.75 & 20.14 & 29.11 & 27.76 & 0.67 & 29.07 & 1.43 \\
  true &  4.41 & 54.19 & 52.42 & 47.70 & 4.72 & 47.70 & 4.72 \\
  relr &  0.17 & 0.37 & 0.56 & 0.58 & 0.14 & 0.61 & 0.30 \\
  \hline
  & \multicolumn{7}{c}{$N$=4,000} \\
  \hline
  bias &  0.34 & 20.02 & 28.98 & 27.65 & 0.02 & 28.96 & 1.33 \\
  sd  & 0.35 & 1.22 & 1.19 & 1.22 & 0.35 & 1.24 & 0.23 \\
  rmse &  0.49 & 20.06 & 29.01 & 27.68 & 0.35 & 28.99 & 1.35 \\
  true &  4.40 & 54.18 & 52.45 & 47.73 & 4.72 & 47.73 & 4.72 \\
  relr &  0.11 & 0.37 & 0.55 & 0.58 & 0.07 & 0.61 & 0.29 \\
  \hline
\end{tabular}
\end{center}
\par
{\footnotesize Note: `bias', `sd', and `rmse' provide the bias, standard deviation, and root mean squared error of the respective estimator. `true' and `relr' are the respective true effect as well as the root mean squared error relative to the true effect.}
\end{table}

In our final simulation design, we maintain the exponential outcome model but assume $D$ to be selective w.r.t.\ $U$ rather than random. To this end, the treatment model in \eqref{sim} is replaced by $D=I\{U+Q>0\}$, with the independent variable $Q\sim N(0,1)$ being an unobserved term. Under this violation of Assumption 7, complier shares and effects are no longer identified, which is confirmed by the simulation results presented in Table \ref{simnonlinnonran}. The bias in the change-in-changes based total, direct, and indirect effects on compliers do not vanish as the sample size increases. Furthermore, under non-random assignment of $D$ (while maintaining monotonicity of $M$ in $D$), the never-takers' and always-takers' respective distributions of $U$ differ across treatment. Therefore, average direct effects among the total of never or always-takers, respectively, are not identified. Yet, $\theta_1^{1,0}$, which is still identified by the same estimator as before, yields the direct effect among never-takers with $D=1$ (as defiers do not exist). Likewise, $\theta_1^{0,1}$ corresponds to the direct effect on always-takers with $D=0$. Indeed, the results in Table \ref{simnonlinnonran} suggest that both parameters are consistently estimated with the change-in-changes model (Panel A.).

\section{Application}\label{app}

Our empirical application is based on the JOBS II data by \cite{VinokurPrice1999}. JOBS II was a randomized job training intervention in the US, designed to analyse the impact of job training on labour market and mental health outcomes, see \cite{VinokurPriceSchul1995}. It was a modified version of the earlier JOBS programme, which had been found to improve labour market outcomes such as job satisfaction, motivation, earnings, and job stability, see \cite{Caplanetal1989} and \cite{VinokurRynGramlichPrice1991}, as well as mental health, see \cite{VinokurPriceCaplan1991}. According to the results of \cite{VinokurPriceSchul1995}, the JOBS II programme increased reemployment rates and improved mental health outcomes, especially for participants having an elevated risk of depression. The JOBS interventions had an important impact in the academic literature \citep[see e.g.][]{Wanberg2012,Liuetal2014} and the methodology was implemented in field experiments in Finland \citep[][]{Vuorietal2002,Vuorietal2005} and the Netherlands \citep[][]{Brenninkmeijer2011}, suggesting positive effects on labour market integration in either case.


The JOBS II intervention  was conducted in south-eastern Michigan, where 2,464 job seekers were eligible to participate in a randomized field experiment, see \cite{VinokurPrice1999}. In a baseline period prior to programme assignment, individuals responded to a screening questionnaire that collected pre-treatment information on mental health. Based on the latter, individuals were classified as having either a high or low depression risk and those with a high risk were oversampled before the training was randomly assigned.\footnote{In the JOBS II intervention, randomization was followed by yet another questionnaire sent out two weeks before the actual job training, see \cite{VinokurPriceSchul1995}, which also provided information on whether an individual had been assigned the training. Consequently, the data collected in that questionnaire must be considered post-treatment as they could be affected by learning the assignment.  Therefore, we rely on the earlier screening data as the relevant pre-treatment period prior to random programme assignment.} The job training consisted of five 4-hours seminars conducted in morning sessions during one week between March 1 and August 7, 1991. Members of the treatment group who participated in at least four of the five sessions received USD 20. Each of the standardized training sessions consisted, among other aspects, of the learning and practicing of job search and problem-solving skills.\footnote{When compared to the earlier JOBS programme \citep[][]{Caplanetal1989}, the job training sessions of JOBS II focused more strongly on building a sense of mastery, personal control and self- efficacy in job search. Previous research had suggested that an increase in this sense of mastery, control and self-efficacy improved observed effort in job search behaviour \citep[][]{EdenAviram1993}. Results in \cite{MarshallLang1990} suggest that mastery is a strong predictor of depression symptoms among women. For a detailed discussion of the literature, the exact sampling process, the training programme, and further aspects, see \cite{VinokurPriceSchul1995}.}  The control group received a booklet with information on job search methods \citep[][p. 44-49]{VinokurPriceSchul1995}.

We analyse the impact of job training on mental health, namely symptoms of depression 6 months after training participation. The health outcome ($Y$) is based on a 11-items index of depression symptoms of the Hopkins Symptom Checklist. For example, respondents were asked how much they were bothered by symptoms such as crying easily, feeling lonely, feeling blue, feeling hopeless, having thoughts of ending their lives, or experiencing a loss of sexual interest. The questions were coded on a 5-point scale, going from `not at all' (1) to `extremely' (5), and summarized in a depression variable that consists of the average across all questions.

One-sided non-compliance with the random assignment is a major issue in JOBS II. While the study design rules out always-takers because members of the control group did not have access to the job training programme, 45\% of those assigned to training in our data did not participate and are therefore never-takers, the remaining 55\% are compliers. In order to avoid selection bias w.r.t\ actual participation, the original JOBS II study by \cite{VinokurPriceSchul1995} analysed the total effect of the policy (i.e.\ the intention-to-treat effect), including those who, despite receiving an offer to participate, did not take part in the job training. In contrast, we use our methodology to separate the direct effect of mere training assignment, which is our treatment $D$, from the indirect effect operating through actual training participation, which is our mediator $M$, among compliers.\footnote{\cite{ImKeTi09} analyse Jobs II in a mediation context as well, but consider a different mediator, namely job search self-efficacy, and a different identification strategy based on selection on observables.}  We also consider the direct effect on never-takers, which likely differs from that on the compliers. While being offered (or not offered) the job training might affect compliers' mental health by inducing motivation/enthusiasm (or discouragement), it may not have the same effect among never-takers, who do not attend such seminars whatsoever.

More concisely, we base identification on Theorem 3a with Assumption 4a for the average direct effect on never-takers, $\theta_1^n$, on Theorem 3b with Assumption 4 for the direct effect on compliers under $d = 0$, $\theta_1^c(0)$, and Theorem 5 with Assumptions 4b and 6a for the indirect effect on compliers under $d = 1$, $\delta_1^c(1)$. None of these approaches requires the presence of always-takers in the sample. We also note that if random assignment operated through other mechanisms than actual participation in any of the subpopulations as it may appear reasonable in the context of mental health outcomes, this would violate the exclusion restriction when using assignment as instrumental variable for actual participation in a two stage least squares regression. Given that our identifying assumptions hold, our approach can therefore be used to statistically test the exclusion restriction.

\begin{table}[]
\caption{Descriptive statistics on outcomes in pre- and post-mediator periods} \label{desc}
\begin{center}
\begin{tabular}{l|cc|cc} \hline\hline
							& \multicolumn{2}{c|}{pre-treatment ($T=0$)} & \multicolumn{2}{c}{post-mediator ($T=1$)} \\
							& sample size     	& mean         	  	& sample size    	& mean   	   \\\hline
overall              	& 1,796  	& 1.86       	     	& 1,564       	& 1.73       	      	 \\                                      			&  	&        	 (0.58)      	 	&        	& (0.67)     	  \\ \hline
$D=0$                                      			& 551    	& 1.87       	      	& 486        	& 1.78      	      	 \\    	&     	& (0.59)     	&       	& (0.70)     	 \\

$D=1$                                          			& 1,245 	& 1.86       	      & 1,078       	& 1.70     	   	 \\
	& 	& (0.57)    	&       	& (0.66)    	  \\ \hline
mean diff                              		&        		& 0.01       	         	&            	&         0.08  	        		 \\
pval                                      			&        		& 0.74      	            	&
	      		&      0.03 		 \\
SD				&        		& 1.63       	            	&            	&    11.79   	 \\             \hline
\end{tabular}
\end{center}
\par
{\footnotesize Note: Standard deviations are in parentheses. `mean diff', `pval', and `SD' are the mean difference, its p-value, and the standardized difference, respectively.}
\end{table}

Our evaluation sample consists of a total of 3,360 observations in the pre-treatment and post-mediator periods with non-missing information for $D$, $M$, and $Y$. It is an unbalanced panel due to attrition of roughly 13\% of the initial respondents between the two periods. Table \ref{desc} provides summary statistics for the outcome in the total sample as well as by treatment group over time. We verify whether randomization was successful by comparing the outcome means of the treatment and control groups in the pre-treatment period ($T=0$) just prior to the randomization of $D$. The small difference of 0.01 is not statistically significant according to a two sample t-test. Furthermore, the standardized difference test suggested by \cite{RosenbaumRubin1985} yields a value of just 1.68 and is thus far below 20, a threshold frequently chosen for indicating problematic imbalances across treatment groups.  To test for potential attrition bias we also consider these statistics in the pre-treatment period exclusively among the panel cases that remain in the sample in the post-mediator period (not reported in Table \ref{desc}). The p-value of the t-test  amounts to 0.52 and the standardized difference of 3.5 is low such that attrition bias does not appear to be a concern. We therefore do not find statistical evidence for a violation of the random assignment of $D$ in our sample. Table \ref{desc} also reports the mean difference in outcomes in the post-mediator period ($T=1$) 6 months after participation, which is an estimate for the total (or intention-to-treat) effect of $D$. The difference of 0.08 is statistically significant at the 5\% level.

\begin{table}[!ht]
	\caption{\label{appl} Empirical results for Jobs II}
	\begin{center}
		\begin{tabular}{r|cccc|cccc|cc}
			\hline\hline
			& \multicolumn{4}{c|}{Changes-in-Changes }  & \multicolumn{4}{c|}{Difference-in-Differences} & \multicolumn{2}{c}{Type shares} \\
			& $\hat{\theta}_1^n$ & $\hat{\Delta}_c$ &  $\hat{\theta}_1^c(0)$ & $\hat{\delta}_1^c(1)$ &  $\hat{\theta}_1^n$ & $\hat{\Delta}_c$ &  $\hat{\theta}_1^c(0)$ & $\hat{\delta}_1^c(1)$ & $\hat{p}(n)$ &  $\hat{p}(c)$\\
			\hline
		est & -0.04 & -0.11 & 0.06 & -0.17 & -0.03 & -0.12 & -0.06 & -0.06 & 0.45 & 0.55 \\
		se & 0.05 & 0.06 & 0.05 & 0.08 & 0.05 & 0.06 & 0.05 & 0.07 & 0.01 & 0.01 \\
		pval & 0.40 & 0.06 & 0.26 & 0.04  & 0.52 & 0.03 & 0.21 & 0.43 & 0.00 & 0.00 \\
			\hline
\end{tabular}
\end{center}
\par
{\footnotesize Note: `est', `se', and `pval' provide the effect estimate, standard error, and p-value of the respective estimator. $\hat{p}(n)$ and  $\hat{p}(c)$ are the estimated never-taker and complier shares. Standard errors are based on cluster bootstrapping the effects 1999 times where clustering is on the respondent level.}
\end{table}

Table \ref{appl} presents the estimation results based on our CiC approach and the DiD strategy of \cite{Deuchertetal2017} when (linearly) controlling for the gender of respondents in either case. Standard errors rely on cluster bootstrapping the direct and indirect effects 1999 times, where clustering is on the respondent level. The CiC and DiD estimates of the direct effects on never-takers, $\hat{\theta}_1^n(0)$, as well as on compliers, $\hat{\theta}_1^c(0)$, are not statistically significant at conventional levels. Hence, we do not find statistical evidence for a direct effect of the mere assignment into the training programme on the depression outcome, which would point to a violation of the exclusion restriction when using assignment as instrument for participation. In contrast, we find for both CiC and DiD negative total effects among compliers $\hat{\Delta}_c$ that are statistically significant at least at the 10\% level. In the case of CiC, also the negative indirect effect among compliers, $\hat{\delta}_1^c(1)$, is significant at the 5\% level, while this is not in the case for DiD.  By and large, our results point to a moderately negative treatment effect on depressive symptoms through actual programme participation, rather than through other (i.e.\ direct) mechanisms. The CiC estimates $\hat{\delta}_1^c(1)$ and $\hat{\Delta}_c$ are in fact rather similar to the result of a two stage least squares regression relying on the exclusion restriction by using $D$ as instrument for $M$. The latter approach yields a local average treatment effect on compliers in the post-mediator period of -0.14 with a heteroskedasticity-robust standard error of 0.07 (significant at the 5\% level).

\section{Conclusion}\label{con}

We proposed a novel identification strategy for causal mediation analysis with repeated cross sections or panel data based on changes-in-changes (CiC) assumptions that are related but yet different to \cite{at06} considering total treatment effects. Strict monotonicity of outcomes in unobserved heterogeneity and distributional time invariance of the latter within groups defined on treatment and mediator states are key assumptions for identifying direct effects within these groups. Additionally assuming random treatment assignment and weak monotonicity of the mediator in the treatment permits identifying direct effects on never-takers and always-takers as well as total, direct, and indirect effects on compliers. We also provided a brief simulation study and an empirical application to the Jobs II programme.

\bibliographystyle{econometrica}
\bibliography{bibliothek_neu}

\bigskip

\renewcommand\appendix{\par
   \setcounter{section}{0}%
   \setcounter{subsection}{0}%
   \setcounter{table}{0}%
	\setcounter{figure}{0}%
   \renewcommand\thesection{\Alph{section}}%
   \renewcommand\thetable{\Alph{section}.\arabic{table}}}
	 \renewcommand\thefigure{\Alph{section}.\arabic{subsection}.\arabic{subsubsection}.\arabic{figure}}
\clearpage

\begin{appendix}

\numberwithin{equation}{section}
\noindent \textbf{\LARGE Appendices}


\section{Proof of Theorem 1 \label{appA}}

\subsection{Average direct effect under $\mathbf{d=1}$ conditional on $\mathbf{D=1}$ and $\mathbf{M(1)=0}$}

In the following, we prove that $\theta_1^{1,0}(1)= E[Y_1(1,0)-Y_1(0,0)|D=1,M_i (1)=0]=  E[Y_1-Q_{00}(Y_0)|D=1,M=0]$. Using the observational rule, we obtain $E[Y_1(1,0)|D=1,M(1)=0]=E[Y_1|D=1,M=0]$. Accordingly, we have to show that $E[Y_1(0,0)|D=1,M(1)=0]=E[Q_{00}(Y_0)|D=1,M=0]$ to finish the proof.

Denote the inverse of $h(d,m,t,u)$ by $h^{-1}(d,m,t;y)$, which exists because of the strict monotonicity required in Assumption 1. Under Assumptions 1 and 3a, the conditional potential outcome distribution function equals
\begin{equation} \label{appA_eq1}
\begin{array}{rl}
 F_{Y_t(d,0)|D=1,M=0}(y) & \stackrel{A1}{=} \Pr(h(d,m,t,U) \leq y|D=1,M=0,T=t) ,\\
&= \Pr(U \leq h^{-1}(d,m,t;y)|D=1,M=0,T=t) ,\\
&\stackrel{A3a}{=} \Pr(U \leq h^{-1}(d,m,t;y)|D=1,M=0) ,\\
&= F_{U|10} ( h^{-1}(d,m,t;y)),
\end{array}
\end{equation}
for $d,d' \in \{0,1\}$. We use these quantities in the following.

First, evaluating $F_{Y_1(0,0)|D=1,M=0}(y)$ at $h(0,0,1,u)$ gives
\begin{equation*}
F_{Y_1(0,0)|D=1,M=0}(h(0,0,1,u)) = F_{U|10} ( h^{-1}(0,0,1;h(0,0,1,u)))  =F_{U|10} ( u).
\end{equation*}
Applying $F_{Y_1(0,0)|D=1,M=0}^{-1}(q)$ to both sides, we have
\begin{equation} \label{appA_eq2}
h(0,0,1,u)  =F_{Y_1(0,0)|D=1,M=0}^{-1}(F_{U|10} ( u)).
\end{equation}
Second, for $F_{Y_0(0,0)|D=1,M=0}(y)$ we have
\begin{equation}\label{appA_eq3}
F_{U|D=1,M=0}^{-1} ( F_{Y_0(0,0)|D=1,M=0}(y)) =   h^{-1}(0,0,0;y).
\end{equation}
Combining (\ref{appA_eq2}) and (\ref{appA_eq3}) yields,
\begin{equation} \label{appA_eq4}
h(0,0,1,h^{-1}(0,0,0;y))  =F_{Y_1(0,0)|D=1,M=0}^{-1} \circ F_{Y_0(0,0)|D=1,M=0}(y) .
\end{equation}
Note that $h(0,0,1,h^{-1}(0,0,0;y))$ maps the period 1 (potential) outcome of an individual with the outcome $y$ in period 0 under non-treatment without the mediator. Accordingly, $E[F_{Y_1(0,0)|D=1,M=0}^{-1} \circ F_{Y_0(0,0)|D=1,M=0}(Y_0)|D=1,M=0]= E[Y_1(0,0)|D=1,M=0]$. We can identify $F_{Y_0(0,0)|D=1,M=0}(y)$ under Assumption 2, but we cannot identify $F_{Y_1(0,0)|D=1,M=0}(y)$. However, we show in the following that we can identify the overall quantile-quantile transform $F_{Y_1(0,0)|D=1,M=0}^{-1} \circ F_{Y_0(0,0)|D=1,M=0}(y)$ under the additional Assumption 3b.

Under Assumptions 1 and 3b, the conditional potential outcome distribution function equals
\begin{equation} \label{appA_eq5}
\begin{array}{rl}
 F_{Y_t(d,0)|D=0,M=0}(y) & \stackrel{A1}{=} \Pr(h(d,m,t,U) \leq y|D=0,M=0,T=t) ,\\
&= \Pr(U \leq h^{-1}(d,m,t;y)|D=0,M=0,T=t) ,\\
&\stackrel{A3b}{=} \Pr(U \leq h^{-1}(d,m,t;y)|D=0,M=0) ,\\
&= F_{U|00} ( h^{-1}(d,m,t;y)),
\end{array}
\end{equation}
for $d,d' \in \{0,1\}$. We repeat similar steps as above. First, evaluating $F_{Y_1(0,0)|D=0,M=0}(y)$ at $h(0,0,1,u)$ gives
\begin{equation*}
F_{Y_1(0,0)D=0,M=0}(h(0,0,1,u)) = F_{U|00} ( h^{-1}(0,0,1;h(0,0,1,u)))  =F_{U|00} ( u).
\end{equation*}
Applying $F_{Y_1(0,0)|D=0,M=0}^{-1}(q)$ to both sides, we have
\begin{equation} \label{appA_eq6}
h(0,0,1,u)  =F_{Y_1(0,0)|D=0,M=0}^{-1}(F_{U|00} ( u)).
\end{equation}
Second, for $F_{Y_0(0,0)|D=0,M=0}(y)$ we have
\begin{equation} \label{appA_eq7}
F_{U|00}^{-1} ( F_{Y_0(0,0)|D=0,M=0}(y)) =   h^{-1}(0,0,0;y).
\end{equation}
Combining (\ref{appA_eq6}) and (\ref{appA_eq7}) yields,
\begin{equation} \label{appA_eq8}
h(0,0,1,h^{-1}(0,0,0;y))  =F_{Y_1(0,0)|D=0,M=0}^{-1} \circ F_{Y_0(0,0)|D=0,M=0}(y) .
\end{equation}

The left sides of (\ref{appA_eq4}) and (\ref{appA_eq8}) are equal. In contrast to (\ref{appA_eq4}), (\ref{appA_eq8}) contains only distributions that can be identified from observable data. In particular, $F_{Y_t(0,0)|D=0,M=0}(y)  =\Pr(Y_t(0,0) \leq y|D=0,M=0) =  \Pr(Y_t \leq y|D=0,M=0)$. Accordingly, we can identify $F_{Y_1(0,0)|D=1,M=0}^{-1} \circ F_{Y_0(0,0)|D=1,M=0}(y)$ by $Q_{00} (y) \equiv F_{Y_1|D=0,M=0}^{-1} \circ F_{Y_0|D=0,M=0}(y) $.

Parsing $Y_0$ through $Q_{00}(\cdot)$ in the treated group without mediator gives
\begin{equation}\label{q00} \begin{array}{rl}
& E[Q_{00}(Y_0)|D=1,M=0] \\& \qquad =  E[F_{Y_1|D=0,M=0}^{-1} \circ F_{Y_0|D=0,M=0}(Y_0 )|D=1,M=0], \\
& \qquad =  E[F_{Y_1(0,0)|D=0,M=0}^{-1} \circ F_{Y_0(0,0)|D=0,M=0}(Y_0(1,0))|D=1,M=0], \\
&\qquad \stackrel{A1,A3b}{=}  E[h(0,0,1,h^{-1}(0,0,0; Y_0(1,0)))|D=1,M=0], \\
&\qquad \stackrel{A2}{=} E[h(0,0,1,h^{-1}(0,0,0; Y_0(0,0)))|D=1,M=0], \\
&\qquad \stackrel{A1,A3a}{=}E[F_{Y_1(0,0)|D=1,M=0}^{-1} \circ F_{Y_0(0,0)|D=1,M=0}(Y_0 (0,0))|D=1,M=0],\\
&\qquad =E[Y_1(0,0)|D=1,M=0]=E[Y_1(0,0)|D=1,M(1)=0], \end{array}
\end{equation}
which has data support because of Assumption 4a.

\subsection{Quantile direct effect under $\mathbf{d=1}$ conditional on $\mathbf{D=1}$ and $\mathbf{M(1)=0}$}

In the following, we prove that
\begin{align*}\theta_1^{1,0}(q,1) &= F_{Y_{1}(1,0)|D=1,M(1)=0}^{-1}(q)-F_{Y_{1}(0,0)|D=1,M(1)=0}^{-1}(q),\\&= F_{Y_{1}|D=1,M=0}^{-1}(q)-F_{Q_{00}(Y_{0})|D=1,M=0}^{-1}(q).
\end{align*} For this purpose, we have to show that
\begin{align}
F_{Y_{1}(1,0)|D=1,M(1)=0}(y)  & = F_{Y_{1}|D=1,M=0}(y) \mbox{ and} \label{qq1} \\
F_{Y_{1}(0,0)|D=1,M(1)=0} (y) & =F_{Q_{00}(Y_{0})|D=1,M=0}(y) \label{qq2},
\end{align}
which is sufficient to show that the quantiles are also identified. We can show (\ref{qq1}) using the observational rule $F_{Y_{1}(1,0)|D=1,M(1)=0}(y)= F_{Y_{1}|D=1,M=0}(y)= E[1\{Y_1 \leq y\} |D=1,M=0]$, with $1\{\cdot\}$ being the indicator function.

Using (\ref{q00}), we obtain
\begin{equation} \label{qqq1} \begin{array}{rl}
&F_{Q_{00}(Y_{0})|D=1,M=0}(y)\\ & \qquad = E[1\{Q_{00}(Y_0) \leq y \}|D=1,M=0],\\
&\qquad =E[1\{ F_{Y_1|D=0,M=0}^{-1} \circ F_{Y_0|D=0,M=0}(Y_0 ) \leq y \} |D=1,M=0],\\
&\qquad =E[1\{ Y_1(0,0)\leq y \}|D=1,M=0],\\ &\qquad =F_{Y_{1}(0,0)|D=1,M(1)=0} (y), \end{array}
\end{equation}
which proves (\ref{qq2}).

\subsection{Average direct effect under $\mathbf{d=0}$ conditional on $\mathbf{D=0}$ and $\mathbf{M(0)=0}$}

In the following, we show that $\theta_1^{0,0}(0)= E[Y_1(1,0)-Y_1(0,0)|D=0,M(0)=0]=  E[Q_{10}(Y_0)-Y_1|D=0,M=0]$. Using the observational rule, we obtain $E[Y_1(0,0)|D=0,M(0)=0]=E[Y_1|D=0,M=0]$. Accordingly, we have to show that $E[Y_1(1,0)|D=0,M(0)=0]=E[Q_{10}(Y_0)|D=0,M=0]$ to finish the proof.

First, we use (\ref{appA_eq5}) to evaluate $F_{Y_1(1,0)|D=0,M=0}(y)$ at $h(1,0,1,u)$
\begin{equation*}
F_{Y_1(1,0)|D=0,M=0}(h(1,0,1,u)) = F_{U|10} ( h^{-1}(1,0,1;h(1,0,1,u)))  =F_{U|10} ( u).
\end{equation*}
Applying $F_{Y_1(1,0)|D=0,M=0}^{-1}(q)$ to both sides, we have
\begin{equation} \label{appA_eq2_12}
h(1,0,1,u)  =F_{Y_1(1,0)|D=0,M=0}^{-1}(F_{U|10} ( u)).
\end{equation}
Second, for $F_{Y_0(1,0)|D=0,M=0}(y)$ we have
\begin{equation}\label{appA_eq3_12}
F_{U|10}^{-1} ( F_{Y_0(1,0)|D=0,M=0}(y)) =   h^{-1}(1,0,0;y),
\end{equation}
using (\ref{appA_eq5}). Combining (\ref{appA_eq2_12}) and (\ref{appA_eq3_12}) yields,
\begin{equation} \label{appA_eq4_12}
h(1,0,1,h^{-1}(1,0,0;y))  =F_{Y_1(1,0)|D=0,M=0}^{-1} \circ F_{Y_0(1,0)|D=0,M=0}(y) .
\end{equation}
Note that $h(1,0,1,h^{-1}(1,0,0;y))$ maps the period 1 (potential) outcome of an individual with the outcome $y$ in period 0 under treatment without the mediator. Accordingly, $E[F_{Y_1(1,0)|D=0,M=0}^{-1} \circ F_{Y_0(1,0)|D=0,M=0}(Y_0)|D=0,M=0]= E[Y_1(1,0)|D=1,M=0]$. We can identify $F_{Y_0(1,0)|D=0,M=0}(y)$ under Assumption 2, but we cannot identify $F_{Y_1(1,0)|D=0,M=0}(y)$. However, we show in the following that we can identify the overall quantile-quantile transform $F_{Y_1(1,0)|D=0,M=0}^{-1} \circ F_{Y_0(1,0)|D=0,M=0}(y)$ under the additional Assumption 3a.

First, we use (\ref{appA_eq1}) to evaluate $F_{Y_1(1,0)|D=1,M=0}(y)$ at $h(1,0,1,u)$
\begin{equation*}
F_{Y_1(1,0)|D=10,M=0}(h(1,0,1,u)) = F_{U|10} ( h^{-1}(1,0,1;h(1,0,1,u)))  =F_{U|10} ( u).
\end{equation*}
Applying $F_{Y_1(1,0)|D=1,M=0}^{-1}(q)$ to both sides, we have
\begin{equation} \label{appA_eq5_12}
h(1,0,1,u)  =F_{Y_1(1,0)|D=1,M=0}^{-1}(F_{U|10} ( u)).
\end{equation}
Second, for $F_{Y_0(1,0)|D=0,M=0}(y)$ we have
\begin{equation} \label{appA_eq6_12}
F_{U|10}^{-1} ( F_{Y_0(1,0)|D=1,M=0}(y)) =   h^{-1}(1,0,0;y),
\end{equation}
using (\ref{appA_eq1}). Combining (\ref{appA_eq5_12}) and (\ref{appA_eq6_12}) yields,
\begin{equation} \label{appA_eq7_12}
h(1,0,1,h^{-1}(1,0,0;y))  =F_{Y_1(1,0)|D=1,M=0}^{-1} \circ F_{Y_0(1,0)|D=1,M=0}(y) .
\end{equation}
The left sides of (\ref{appA_eq4_12}) and (\ref{appA_eq7_12}) are equal. In contrast to (\ref{appA_eq4_12}), (\ref{appA_eq7_12}) contains only distributions that can be identified from observable data. In particular, $F_{Y_t(1,0)|D=1,M=0}(y)  =\Pr(Y_t(1,0) \leq y|D=1,M=0) =  \Pr(Y_t \leq y|D=1,M=0)$. Accordingly, we can identify $F_{Y_1(1,0)|D=0,M=0}^{-1} \circ F_{Y_0(1,0)|D=0,M=0}(y)$ by $Q_{10} (y) \equiv F_{Y_1|D=1,M=0}^{-1} \circ F_{Y_0|D=1,M=0}(y) $.

Parsing $Y_0$ through $Q_{10}(\cdot)$ in the non-treated group without mediator gives
\begin{equation} \label{q10} \begin{array}{rl}
&E[Q_{10}(Y_0)|D=0,M=0] \\ & \qquad =  E[F_{Y_1|D=1,M=0}^{-1} \circ F_{Y_0|D=1,M=0}(Y_0 )|D=0,M=0], \\
& \qquad =  E[F_{Y_1(1,0)|D=1,M=0}^{-1} \circ F_{Y_0(1,0)|D=1,M=0}(Y_0(0,0))|D=0,M=0], \\
&\qquad \stackrel{A1,A3a}{=}  E[h(1,0,1,h^{-1}(1,0,0; Y_0(0,0)))|D=0,M=0], \\
&\qquad \stackrel{A2}{=} E[h(1,0,1,h^{-1}(1,0,0; Y_0(1,0)))|D=1,M=0] ,\\
&\qquad \stackrel{A1,A3b}{=}E[F_{Y_1(1,0)|D=0,M=0}^{-1} \circ F_{Y_0(1,0)|D=0,M=0}(Y_0 (1,0))|D=0,M=0],\\
&\qquad =E[Y_1(1,0)|D=0,M=0]=E[Y_1(1,0)|D=0,M(0)=0],
\end{array} \end{equation}
which has data support because of Assumption 4b.

\subsection{Quantile direct effect under $\mathbf{d=0}$ conditional on $\mathbf{D=0}$ and $\mathbf{M(0)=0}$}

In the following, we prove that
\begin{align*}
\theta_1^{0,0}(q,0) &= F_{Y_{1}(1,0)|D=0,M(0)=0}^{-1}(q)-F_{Y_{1}(0,0)|D=0,M(0)=0}^{-1}(q),\\&= F_{Q_{10}(Y_{0})|D=0,M=0}^{-1}(q)-F_{Y_{1}|D=0,M=0}^{-1}(q).
\end{align*}
For this purpose, we have to show that
\begin{align}
F_{Y_{1}(1,0)|D=0,M(0)=0}(y)  & = F_{Q_{10}(Y_{0})|D=0,M=0}(y) \mbox{ and} \label{qq3} \\
F_{Y_{1}(0,0)|D=0,M(0)=0}(y) & =F_{Y_{1}|D=0,M=0}(y) \label{qq4},
\end{align}
which is sufficient to show that the quantiles are also identified. We can show (\ref{qq4}) using the observational rule $F_{Y_{1}(0,0)|D=0,M(0)=0}(y)= F_{Y_{1}|D=0,M=0}(y)= E[1\{Y_1 \leq y\} |D=0,M=0]$.

Using (\ref{q10}), we obtain
\begin{equation*} \begin{array}{rl}
&F_{Q_{10}(Y_{0})|D=0,M=0}(y) \\ & \qquad = E[1\{Q_{10}(Y_0) \leq y \}|D=0,M=0],\\
&\qquad =E[1\{ F_{Y_1|D=1,M=0}^{-1} \circ F_{Y_0|D=1,M=0}(Y_0 ) \leq y \} |D=0,M=0],\\
&\qquad =E[1\{ Y_1(1,0)\leq y \}|D=0,M=0],\\
&\qquad =F_{Y_{1}(1,0)|D=0,M(0)=0}(y), \end{array}
\end{equation*}
which proves (\ref{qq3}).

\section{Proof of Theorem 2 \label{appB}}

\subsection{Average direct effect under $\mathbf{d=0}$ conditional on $\mathbf{D=0}$ and $\mathbf{M(0)=1}$}

In the following, we show that $\theta_1^{0,1}(0)= E[Y_1(1,1)-Y_1(0,1)|D=0,M(0)=1]=  E[Q_{11}(Y_0)-Y_1|D=0,M=1]$. Using the observational rule, we obtain $E[Y_1(0,1)|D=0,M(0)=1]=E[Y_1|D=0,M=1]$. Accordingly, we have to show that $E[Y_1(1,1)|D=0,M(0)=1]=E[Q_{11}(Y_0)|D=0,M=1]$ to finish the proof.

Under Assumptions 1 and 5a, the conditional potential outcome distribution function equals
\begin{equation} \label{appA_eq1_2}
\begin{array}{rl}
 F_{Y_t(d,0)|D=1,M=0}(y) & \stackrel{A1}{=} \Pr(h(d,m,t,U) \leq y|D=0,M=1,T=t) ,\\
&= \Pr(U \leq h^{-1}(d,m,t;y)|D=0,M=1,T=t) ,\\
&\stackrel{A5a}{=} \Pr(U \leq h^{-1}(d,m,t;y)|D=0,M=1) ,\\
&= F_{U|01} ( h^{-1}(d,m,t;y)),
\end{array}
\end{equation}
for $d,d' \in \{0,1\}$. We use these quantities in the following.

First, evaluating $F_{Y_1(1,1)|D=0,M=1}(y)$ at $h(1,1,1,u)$ gives
\begin{equation*}
F_{Y_1(1,1)|D=0,M=1}(h(1,1,1,u)) = F_{U|01} ( h^{-1}(1,1,1;h(1,1,1,u)))  =F_{U|01} ( u).
\end{equation*}
Applying $F_{Y_1(1,1)|D=0,M=1}^{-1}(q)$ to both sides, we have
\begin{equation} \label{appA_eq2_2}
h(1,1,1,u)  =F_{Y_1(1,1)|D=0,M=1}^{-1}(F_{U|01} ( u)).
\end{equation}
Second, for $F_{Y_0(1,1)|D=0,M=1}(y)$ we have
\begin{equation}\label{appA_eq3_2}
F_{U|01}^{-1} ( F_{Y_0(1,1)|D=0,M=1}(y)) =   h^{-1}(1,1,0;y).
\end{equation}
Combining (\ref{appA_eq2_2}) and (\ref{appA_eq3_2}) yields,
\begin{equation} \label{appA_eq4_2}
h(1,1,1,h^{-1}(1,1,0;y))  =F_{Y_1(1,1)|D=0,M=1}^{-1} \circ F_{Y_0(1,1)|D=0,M=1}(y) .
\end{equation}
Note that $h(1,1,1,h^{-1}(1,1,0;y))$ maps the period 1 (potential) outcome of an individual with the outcome $y$ in period 0 under treatment with the mediator. Accordingly, $E[F_{Y_1(1,1)|D=0,M=1}^{-1} \circ F_{Y_0(1,1)|D=0,M=1}(Y_0)|D=0,M=1]= E[Y_1(1,1)|D=0,M=1]$. We can identify $F_{Y_0(1,1)|D=0,M=1}(y)= F_{Y_0|D=0,M=1}(y)$ under Assumption 2, but we cannot identify $F_{Y_1(1,1)|D=0,M=1}(y)$. However, we show in the following that we can identify the overall quantile-quantile transform $F_{Y_1(1,1)|D=0,M=1}^{-1} \circ F_{Y_0(1,1)|D=0,M=1}(y)$ under the additional Assumption 5b.

Under Assumptions 1 and 5b, the conditional potential outcome distribution function equals
\begin{equation} \label{appA_eq5_2}
\begin{array}{rl}
 F_{Y_t(d,1)|D=1,M=1}(y) & \stackrel{A1}{=} \Pr(h(d,m,t,U) \leq y|D=1,M=1,T=t) ,\\
&= \Pr(U \leq h^{-1}(d,m,t;y)|D=1,M=1,T=t) ,\\
&\stackrel{A5b}{=} \Pr(U \leq h^{-1}(d,m,t;y)|D=1,M=1) ,\\
&= F_{U|11} ( h^{-1}(d,m,t;y)),
\end{array}
\end{equation}
for $d,d' \in \{0,1\}$. We repeat similar steps as above. First, evaluating $F_{Y_1(1,1)|D=1,M=1}(y)$ at $h(1,1,1,u)$ gives
\begin{equation*}
F_{Y_1(1,1)|D=1,M=1}(h(1,1,1,u)) = F_{U|11} ( h^{-1}(1,1,1;h(1,1,1,u)))  =F_{U|11} ( u).
\end{equation*}
Applying $F_{Y_1(1,1)|D=1,M=1}^{-1}(q)$ to both sides, we have
\begin{equation} \label{appA_eq6_2}
h(1,1,1,u)  =F_{Y_1(1,1)|D=1,M=1}^{-1}(F_{U|11} ( u)).
\end{equation}
Second, for $F_{Y_0(1,1)|D=1,M=1}(y)$ we have
\begin{equation} \label{appA_eq7_2}
F_{U|11}^{-1} ( F_{Y_0(1,1)|D=1,M=1}(y)) =   h^{-1}(1,1,1;y).
\end{equation}
Combining (\ref{appA_eq6_2}) and (\ref{appA_eq7_2}) yields,
\begin{equation} \label{appA_eq8_2}
h(1,1,1,h^{-1}(1,1,0;y))  =F_{Y_1(1,1)|D=1,M=1}^{-1} \circ F_{Y_0(1,1)|D=1,M=1}(y) .
\end{equation}

The left sides of (\ref{appA_eq4_2}) and (\ref{appA_eq8_2}) are equal. In contrast to (\ref{appA_eq4_2}), (\ref{appA_eq8_2}) contains only distributions that can be identified from observable data. In particular, $F_{Y_t(1,1)|D=1,M=1}(y)  =\Pr(Y_t(1,1) \leq y|D=1,M=1) =  \Pr(Y_t \leq y|D=1,M=1)$. Accordingly, we can identify $F_{Y_1(1,1)|D=0,M=1}^{-1} \circ F_{Y_0(1,1)|D=0,M=1}(y)$ by $Q_{11} (y) \equiv F_{Y_1|D=1,M=1}^{-1} \circ F_{Y_0|D=1,M=1}(y) $.

Parsing $Y_0$ through $Q_{11}(\cdot)$ in the non-treated group with mediator gives
\begin{equation} \label{q11} \begin{array}{rl}
& E[Q_{11}(Y_0)|D=0,M=1]\\ &  \qquad =  E[F_{Y_1|D=1,M=1}^{-1} \circ F_{Y_0|D=1,M=1}(Y_0 )|D=0,M=1], \\
& \qquad  =  E[F_{Y_1(1,1)|D=1,M=1}^{-1} \circ F_{Y_0(1,1)|D=1,M=1}(Y_0(0,1))|D=0,M=1], \\
& \qquad \stackrel{A1,A5b}{=}  E[h(1,1,1,h^{-1}(1,1,0; Y_0(0,1)))|D=0,M=1], \\
& \qquad \stackrel{A2}{=} E[h(1,1,1,h^{-1}(1,1,0; Y_0(0,0)))|D=0,M=1], \\
& \qquad \stackrel{A1,A5a}{=}E[F_{Y_1(1,1)|D=0,M=1}^{-1} \circ F_{Y_0(1,1)|D=0,M=1}(Y_0 (0,0))|D=0,M=1],\\
& \qquad =E[Y_1(1,1)|D=0,M=1]=E[Y_1(1,1)|D=0,M(0)=1],
\end{array} \end{equation}
which has data support because of Assumption 6a.

\subsection{Quantile direct effect under $\mathbf{d=0}$ conditional on $\mathbf{D=0}$ and $\mathbf{M(0)=1}$}

In the following, we show that
\begin{align*}
\theta_1^{0,1}(q,0)&= F_{Y_{1}(1,1)|D=0,M(0)=1}^{-1}(q)-F_{Y_{1}(0,1)|D=0,M(0)=1}^{-1}(q),\\
&= F_{Q_{11}(Y_{0})|D=0,M=1}^{-1}(q)-F_{Y_{1}|D=0,M=1}^{-1}(q).
\end{align*}
 For this purpose, we have to prove that
\begin{align}
 F_{Y_{1}(1,1)|D=0,M(0)=1}(y)  & = F_{Q_{11}(Y_{0})|D=0,M=1}(y) \mbox{ and} \label{qq5} \\
F_{Y_{1}(0,1)|D=0,M(0)=1}(y) & =F_{Y_{1}|D=0,M=1}(y) \label{qq6},
\end{align}
which is sufficient to show that the quantiles are also identified. We can show (\ref{qq6}) using the observational rule $F_{Y_{1}(0,1)|D=0,M(0)=1}(y)= F_{Y_{1}|D=0,M=1}(y)= E[1\{Y_1 \leq y\} |D=0,M=1]$.

Using (\ref{q11}), we obtain
\begin{equation}\label{qqq2} \begin{array}{rl}
&F_{Q_{11}(Y_{0})|D=0,M=1}(y)\\ & \qquad = E[1\{Q_{11}(Y_0) \leq y \}|D=0,M=1],\\
& \qquad =E[1\{F_{Y_1|D=1,M=1}^{-1} \circ F_{Y_0|D=1,M=1}(Y_0 ) \leq y \} |D=0,M=1],\\
& \qquad =E[1\{ Y_1(1,1)\leq y \}|D=0,M=0],\\
& \qquad  = F_{Y_{1}(1,1)|D=0,M(0)=1}(y), \end{array}
\end{equation}
which proves (\ref{qq5}).

\subsection{Average direct effect under $\mathbf{d=1}$ conditional on $\mathbf{D=1}$ and $\mathbf{M(1)=1}$}

In the following, we show that $\theta_1^{1,1}(1)= E[Y_1(1,1)-Y_1(0,1)|D=1,M(1)=1]=  E[Y_1-Q_{01}(Y_0)|D=1,M=1]$. Using the observational rule, we obtain $E[Y_1(1,1)|D=1,M(1)=1]=E[Y_1|D=1,M=1]$. Accordingly, we have to show that $E[Y_1(0,1)|D=1,M(1)=1]=E[Q_{01}(Y_0)|D=1,M=1]$ to finish the proof.

First, using (\ref{appA_eq5_2}) to evaluate $F_{Y_1(0,1)|D=1,M=1}(y)$ at $h(0,1,1,u)$ gives
\begin{equation*}
F_{Y_1(0,1)|D=1,M=1}(h(0,1,1,u)) = F_{U|11} ( h^{-1}(0,1,1;h(0,1,1,u)))  =F_{U|11} ( u).
\end{equation*}
Applying $F_{Y_1(0,1)|D=1,M=1}^{-1}(q)$ to both sides, we have
\begin{equation} \label{appA_eq2_22}
h(0,1,1,u)  =F_{Y_1(0,1)|D=1,M=1}^{-1}(F_{U|11} ( u)).
\end{equation}
Second, for $F_{Y_0(0,1)|D=0,M=1}(y)$ we obtain
\begin{equation}\label{appA_eq3_22}
F_{U|11}^{-1} ( F_{Y_0(0,1)|D=1,M=1}(y)) =   h^{-1}(0,1,0;y),
\end{equation}
using (\ref{appA_eq5_2}). Combining (\ref{appA_eq2_22}) and (\ref{appA_eq3_22}) yields,
\begin{equation} \label{appA_eq4_22}
h(0,1,1,h^{-1}(0,1,0;y))  =F_{Y_1(0,1)|D=1,M=1}^{-1} \circ F_{Y_0(0,1)|D=1,M=1}(y) .
\end{equation}
Note that $h(0,1,1,h^{-1}(0,1,0;y))$ maps the period 1 (potential) outcome of an individual with the outcome $y$ in period 0 under non-treatment with the mediator. Accordingly, $E[F_{Y_1(1,1)|D=0,M=1}^{-1} \circ F_{Y_0(1,1)|D=0,M=1}(Y_0)|D=0,M=1]= E[Y_1(1,1)|D=0,M=1]$. We can identify $F_{Y_0(1,1)|D=0,M=1}(y)= F_{Y_0|D=0,M=1}(y)$ under Assumption 2, but we cannot identify $F_{Y_1(1,1)|D=0,M=1}(y)$. However, we show in the following that we can identify the overall quantile-quantile transform $F_{Y_1(1,1)|D=0,M=1}^{-1} \circ F_{Y_0(1,1)|D=0,M=1}(y)$ under the additional Assumption 5a.

First, using (\ref{appA_eq1_2}) to evaluate $F_{Y_1(0,1)|D=0,M=1}(y)$ at $h(0,1,1,u)$ gives
\begin{equation*}
F_{Y_1(0,1)|D=0,M=1}(h(0,1,1,u)) = F_{U|01} ( h^{-1}(0,1,1;h(0,1,1,u)))  =F_{U|01} ( u).
\end{equation*}
Applying $F_{Y_1(0,1)|D=0,M=1}^{-1}(q)$ to both sides, we have
\begin{equation} \label{appA_eq6_22}
h(0,1,1,u)  =F_{Y_1(0,1)|D=0,M=1}^{-1}(F_{U|01} ( u)).
\end{equation}
Second, for $F_{Y_0(0,1)|D=0,M=1}(y)$ we obtain
\begin{equation} \label{appA_eq7_22}
F_{U|01}^{-1} ( F_{Y_0(0,1)|D=0,M=1}(y)) =   h^{-1}(0,1,1;y),
\end{equation}
using (\ref{appA_eq1_2}). Combining (\ref{appA_eq6_22}) and (\ref{appA_eq7_22}) yields,
\begin{equation} \label{appA_eq8_22}
h(0,1,1,h^{-1}(0,1,0;y))  =F_{Y_1(0,1)|D=0,M=1}^{-1} \circ F_{Y_0(0,1)|D=0,M=1}(y) .
\end{equation}

The left sides of (\ref{appA_eq4_22}) and (\ref{appA_eq8_22}) are equal. In contrast to (\ref{appA_eq4_22}), (\ref{appA_eq8_22}) contains only distributions that can be identified from observable data. In particular, $F_{Y_t(0,1)|D=0,M=1}(y)  =\Pr(Y_t(0,1) \leq y|D=0,M=1) =  \Pr(Y_t \leq y|D=0,M=1)$. Accordingly, we can identify $F_{Y_1(0,1)|D=1,M=1}^{-1} \circ F_{Y_0(0,1)|D=1,M=1}(y)$ by $Q_{01} (y) \equiv F_{Y_1|D=0,M=1}^{-1} \circ F_{Y_0|D=0,M=1}(y) $.

Parsing $Y_0$ through $Q_{01}(\cdot)$ in the treated group with mediator gives
\begin{equation} \label{q01} \begin{array}{rl}
& E[Q_{01}(Y_0)|D=1,M=1] \\& \qquad =  E[F_{Y_1|D=0,M=1}^{-1} \circ F_{Y_0|D=0,M=1}(Y_0 )|D=1,M=1] ,\\
& \qquad  =  E[F_{Y_1(0,1)|D=0,M=1}^{-1} \circ F_{Y_0(0,1)|D=0,M=1}(Y_0(1,1))|D=1,M=1] ,\\
& \qquad \stackrel{A1,A5a}{=}  E[h(0,1,1,h^{-1}(0,1,0; Y_0(1,1)))|D=1,M=1], \\
& \qquad \stackrel{A2}{=} E[h(0,1,1,h^{-1}(0,1,0; Y_0(0,1)))|D=1,M=1], \\
& \qquad \stackrel{A1,A5b}{=}E[F_{Y_1(0,1)|D=1,M=1}^{-1} \circ F_{Y_0(0,1)|D=1,M=1}(Y_0 (0,1))|D=1,M=1],\\
& \qquad =E[Y_1(0,1)|D=1,M=1]=E[Y_1(0,1)|D=1,M(1)=1],
\end{array} \end{equation}
which has data support under Assumption 6b.

\subsection{Quantile direct effect under $\mathbf{d=1}$ conditional on $\mathbf{D=1}$ and $\mathbf{M(1)=1}$}

In the following, we show that
\begin{align*}
\theta_1^{1,1}(q,1)&= F_{Y_{1}(1,1)|D=1,M(1)=1}^{-1}(q)-F_{Y_{1}(0,1)|D=1,M(1)=1}^{-1}(q), \\
&= F_{Y_{1}|D=1,M=1}^{-1}(q)-F_{Q_{01}(Y_{0})|D=1,M=1}^{-1}(q).
\end{align*}
For this purpose, we have to prove that
\begin{align}
F_{Y_{1}(1,1)|D=1,M(1)=1}(y)  & = F_{Y_{1}|D=1,M=1}(y) \mbox{ and} \label{qq7} \\
F_{Y_{1}(0,1)|D=1,M(1)=1}(y) & =F_{Q_{01}(Y_{0})|D=1,M=1}(y) \label{qq8},
\end{align}
which is sufficient to show that the quantiles are also identified. We can show (\ref{qq7}) using the observational rule $F_{Y_{1}(1,1)|D=1,M(1)=1}(y)= F_{Y_{1}|D=1,M=1}(y)= E[1\{Y_1 \leq y\} |D=1,M=1]$.

Using (\ref{q01}), we obtain
\begin{equation*} \begin{array}{rl}
&F_{Q_{01}(Y_{0})|D=1,M=1}(y)  \\ & \qquad = E[1\{Q_{01}(Y_0) \leq y \}|D=1,M=1],\\
& \qquad =E[1\{F_{Y_1|D=0,M=1}^{-1} \circ F_{Y_0|D=0,M=1}(Y_0 ) \leq y \} |D=1,M=1],\\
& \qquad =E[1\{ Y_1(0,1)\leq y \}|D=1,M=0], \\
& \qquad = F_{Y_{1}(0,1)|D=1,M(1)=1}(y), \end{array}
\end{equation*}
which proves (\ref{qq8}).

\section{Proof of equations (\ref{eq002a}) and (\ref{eq002c}) \label{appC}}

$\Delta_1= E[Y_1 |D=1]- E[Y_1 |D=0]$ and quantile treatment effect $\Delta_1(q) = F_{Y_{1} |D=1}^{-1}(q)- F_{Y_{1} |D=0}^{-1}(q)$

The average total effect for the entire population is identified by,
\begin{align*}
\Delta_1 &= E[Y_1(1,M(1)) ]- E[Y_1(0,M(0)) ] , \\
&\stackrel{A7}{=} E[Y_1(1,M(1))|D=1 ]- E[Y_1(0,M(0))|D=0 ] , \\
&= E[Y_1|D=1 ]- E[Y_1|D=0 ] ,
\end{align*}
where the first equality is the definition of $\Delta_1$, the second equality hold by Assumption 7, and the last equality holds by the observational rule.

We define the conditional distribution $F_{Y_{1} |D=d}(y)= \Pr(Y_1 \leq y|D=d)$ and  $F_{Y_{1} |D=d}^{-1}(q)= \inf\{y: F_{Y_{1} |D=d}(y) \geq q\}$. We can show the identification of the total QTE for the entire population  $\Delta_1(q) = F_{Y_{1} |D=1}^{-1}(q)- F_{Y_{1} |D=0}^{-1}(q)$ when we show that $F_{Y_{1}(1,M(1))}(y)= F_{Y_{1} |D=1}(y)$ and $F_{Y_{1}(0,M(0))}(y)= F_{Y_{1} |D=0}(y)$. Using Assumption 7 and the observational rule gives,
\begin{align*}
F_{Y_{1}(1,M(1))}(y) &=\Pr(Y_1(1,M(1)) \leq y) , \\
&\stackrel{A7}{=} \Pr(Y_1(1,M(1)) \leq y|D=1) , \\
&= \Pr(Y_1\leq y|D=1) =F_{Y_{1} |D=1}(y),
\end{align*}
and
\begin{align*}
F_{Y_{1}(0,M(0))}(y) &=\Pr(Y_1(0,M(0)) \leq y) , \\
&\stackrel{A7}{=} \Pr(Y_1(0,M(0)) \leq y|D=0) , \\
&= \Pr(Y_1\leq y|D=0) =F_{Y_{1} |D=0}(y),
\end{align*}
which finishes the proof.

By Assumption 7, the share of a type $\tau$ conditional on $D$ corresponds to $p_{\tau}$ (in the population), as $D$ is randomly assigned. This implies that $p_{1|1} = p_a + p_c$, $p_{1|0} = p_a + p_{de}$, $p_{0|1} = p_n + p_{de}$, and $p_{0|0} = p_n + p_c$. Under Assumption 8, $p_{de}=0$, which finishes the proof of (\ref{eq002a}).

Furthermore, $E[Y_t(d,m)|\tau,D=1]=E[Y_t(d,m)|\tau,D=0]=E[Y_t(d,m)|\tau]$ due to the independence of $D$ and the potential outcomes as well as the types $\tau$ (which are a deterministic function of $M(d)$) under Assumption 7. It follows that conditioning on $D$ is not required on the right hand side of the following equation, which expresses the mean outcome conditional $D=0$ and $M=0$ as weighted average of the mean potential outcomes of compliers and never-takers:
\begin{equation}\label{a1}
\begin{array}{rl}
&E[Y_t|D=0,M=0] \\ & \displaystyle \qquad =\frac{p_n}{p_n + p_c}E[Y_t(0,0)|\tau=n] + \frac{p_c}{p_n + p_c}E[Y_t(0,0)|\tau=c].
\end{array}
\end{equation}
Only compliers and never-takers satisfy $M(0)=0$ and thus make up the group with $D=0$ and $M=0$. After some rearrangements we obtain
\begin{equation}\label{a2}
\begin{array}{rl}
&E[Y_t(0,0)|\tau=n]- E[Y_t(0,0)|\tau=c]\\ & \displaystyle \qquad = \frac{p_n+ p_c}{p_c } \left\{ E[Y_t(0,0)|\tau=n]- E[Y_t|D=0,M=0] \right\}.
\end{array}
\end{equation}
Next, we consider observations with $D=1$ and $M=0$, which might consist of both never-takers and defiers, as $M(1)=0$ for both types. However, by Assumption 8, defiers are ruled out, such that the mean outcome given $D_1=1$ and $M_1=0$ is determined by never-takers only:
\begin{equation}
E[Y_t|D=1,M=0] \stackrel{A7,A8}{=}E[Y_t(1,0)|\tau=n] . \label{a3}
\end{equation}
Furthermore, by Assumption 2,
\begin{equation*}
E[Y_0(0,0)|\tau=n] \stackrel{A2}{=}E[Y_0(1,0)|\tau=n] \stackrel{A7,A8}{=} E[Y_0|D=1,M=0].
\end{equation*}

Similarly to (\ref{a1}) for the never-takers and compliers, consider the mean outcome given $Z=1$ and $D=1$, which is made up by always-takers and compliers (the types with $M(1)=1$)
\begin{equation} \label{a1b}
\begin{array}{rl}
&E[Y_t|D=1,M=1] \\ & \displaystyle \qquad=\frac{p_a}{p_a + p_c}E[Y_t(1,1)|\tau=a] + \frac{p_c}{p_a + p_c}E[Y_t(1,1)|\tau=c].
\end{array}
\end{equation}
After some rearrangements we obtain
\begin{equation} \label{a14}
\begin{array}{rl}
& E[Y_t(1,1)|\tau=a]- E[Y_t(1,1)|\tau=c] \\ & \displaystyle \qquad= \frac{p_a+ p_c}{p_c } \left\{ E[Y_t(1,1)|\tau=a]- E[Y_t|D=1,M=1] \right\}.
\end{array}
\end{equation}
By Assumptions 7 and 8,
\begin{equation} \label{a15}
E[Y_t|D=0,M=1] =E[Y_t(0,1)|\tau=a].
\end{equation}
Now consider (\ref{a14}) for period $T=0$, and note that by Assumption 2, $E[Y_0(1,1)|\tau=a]=E[Y_0(0,0)|\tau=a]=E[Y_0(0,1)|\tau=a]$ and $E[Y_0(1,1)|\tau=c]=E[Y_0(0,0)|\tau=c]$.

Combining (\ref{a1b}), (\ref{a15}), and the law of iterative expectations (LIE) gives
\begin{align*}
&E[Y_0|D=1] \\ & \qquad \stackrel{LIE}{=} E[Y_0|D=1, M=1] \cdot p_{1|1} + E[Y_0|D=1, M=0] \cdot p_{0|1}, \\
&  \qquad = E[Y_0(1,1)|\tau=c] \cdot  p_c + E[Y_0(1,1)|\tau=a]  \cdot p_a + E[Y_0(1,0)|\tau=n]  \cdot p_n, \\
&  \qquad \stackrel{A2}{=} E[Y_0(1,1)|\tau=c]  \cdot p_c + E[Y_0(1,1)|\tau=a] \cdot  p_a + E[Y_0(0,0)|\tau=n]  \cdot p_n.
\end{align*}
Likewise, combining (\ref{a1}) and (\ref{a3}) gives
\begin{align*}
&E[Y_0|D=0] \\&  \qquad \stackrel{LIE}{=} E[Y_0|D=0, M=1] \cdot p_{1|0} + E[Y_0|D_0=1, M=0] \cdot p_{0|0}, \\
&  \qquad = E[Y_0(0,1)|\tau=a] \cdot  p_a + E[Y_0(0,0)|\tau=c]  \cdot p_c +  E[Y_0(0,0)|\tau=n] \cdot  p_n, \\
&  \qquad \stackrel{A2}{=} E[Y_0(1,1)|\tau=a]  \cdot p_a + E[Y_0(0,0)|\tau=c] \cdot  p_c +  E[Y_0(0,0)|\tau=n]  \cdot p_n.
\end{align*}
Accordingly,
\begin{equation*}
 \frac{E[Y_0|D=1] -E[Y_0|D=0]}{p_{1|1} - p_{1|0}}   = E[Y_0(1,1)|\tau=c] - E[Y_0(0,0)|\tau=c]\stackrel{A2}{=} 0,
\end{equation*}
which proves (\ref{eq002c}). Accordingly, $E[Y_0|D=1] -E[Y_0|D=0]=0$ is a testable implication of Assumption 2, 7, and 8.

\section{Proof of Theorem 3 \label{appD}}
\subsection{Average direct effect on the never-takers}

In the following, we show that $\theta_1^n= E[Y_1(1,0)-Y_1(0,0)|\tau=n]= E[Y_1-Q_{00}(Y_0)|D=1,M=0]$. From (\ref{a3}), we obtain the first ingredient $E[Y_1(1,0)|\tau=n]=E[Y_1|D=1,M=0]$. Furthermore, from (\ref{q00}) we have $E[Q_{00}(Y_0)|D=1,M=0] = E[Y_1(0,0)|D=1,M(1)=0]$. Under Assumption 7 and 8,
\begin{equation} \label{pp1} \begin{array}{rl} E[Y_1(0,0)|D=1,M(1)=0]&\stackrel{A7}{=}E[Y_1(0,0)|D=1,\tau=n]\\&\stackrel{A8}{=}E[Y_1(0,0)|\tau=n].
\end{array} \end{equation}

\subsection{Quantile direct effect on the never-takers}

We prove that
\begin{align*}
\theta_1^n (q)&= F_{Y_{1}(1,0)|\tau=n}^{-1}(q)- F_{Y_{1}(0,0)|\tau=n}^{-1}(q), \\
&= F_{Y_1|D=1,M=0}^{-1}(q)-F_{Q_{00}(Y_{0})|D=1,M=0}^{-1}(q).
\end{align*}
 This requires showing that
\begin{align}
F_{Y_{1}(1,0)|\tau=n}(y) &=F_{Y_1|D=1,M=0}(y) \mbox{ and} \label{qq9}\\
F_{Y_{1}(0,0)|\tau=n}(y) &= F_{Q_{00}(Y_{0})|D=1,M=0}(y). \label{qq10}
\end{align}
Under Assumptions 7 and 8,
\begin{equation} \label{a3b}
\begin{array}{rl}
F_{Y_t|D=1,M=0} (y) &= E[1\{Y_t\leq y\}|D=1,M=0] \\& \stackrel{A7,A8}{=}E[1\{Y_t(1,0)\leq y\}|\tau=n] \\& = F_{Y_{t}(1,0)|\tau=n} (y), \end{array}
\end{equation}
which proves (\ref{qq9}). From (\ref{qqq1}), we have
\begin{equation*}
F_{Q_{00}(Y_{0})|D=1,M=0}(y) = F_{Y_{1}(0,0)|D=1,M(1)=0} (y) = E[1\{Y_1(0,0) \leq y\}|D=1,M(1)=0].
\end{equation*}
Under Assumption 7 and 8,
\begin{equation} \label{qqq3} \begin{array}{rl} E[1\{Y_1(0,0) \leq y\}|D=1,M(1)=0]&\stackrel{A7,A8}{=}E[1\{Y_1(0,0) \leq y\}|\tau=n]\\ & = F_{Y_{1}(0,0)|\tau=n}(y),
\end{array} \end{equation}
which proves (\ref{qq10}).

\subsection{Average direct effect under $\mathbf{d = 0}$ on compliers}

In the following, we show that
\begin{align*}
\displaystyle \theta_1^{c}(0) =& E[Y_1(1,0)-Y_1(0,0)|\tau=c], \\ =&
\frac{p_{0|0}}{p_{0|0} - p_{0|1}}E[Q_{10}(Y_0)- Y_1|D=0,M=0] \\
&  - \frac{p_{0|1}}{p_{0|0} - p_{0|1}}E[Y_1 - Q_{00}(Y_0)|D=1,M=0].
\end{align*}
Plugging (\ref{pp1}) in (\ref{a1}) under $T=1$, we obtain
\begin{equation*}
\begin{array}{rl}
E[Y_1|D=0,M=0]  =& \displaystyle  \frac{p_n}{p_n + p_c}E[Q_{00}(Y_0)|D=1,M=0] \\
&\displaystyle \quad +   \frac{p_c}{p_n + p_c}E[Y_1(0,0)|\tau=c].
\end{array}
\end{equation*}
This allows identifying
\begin{equation} \label{p1}
\begin{array}{rl}
E[Y_1(0,0)|\tau=c] =& \displaystyle \frac{p_{0|0}}{ p_{0|0} - p_{0|1}} E[Y_1|D=0,M=0] \\
& \displaystyle \quad -\frac{p_{0|1}}{p_{0|0} - p_{0|1}}E[Q_{00}(Y_0)|D=1,M=0].
\end{array}
\end{equation}

Accordingly, we have to show the identification of $E[Y_1(1,0)|c]$ to finish the proof. From (\ref{q10}) we have $E[Y_1(1,0)|D=0,M=0] = E[Q_{10}(Y_0)|D=0,M=0]$. Applying the law of iterative expectations, gives
\begin{align*}
E[Y_1(1,0)|D=0,M=0] =& \frac{p_n}{p_n+p_c} E[Y_1(1,0)|D=0,M=0, \tau=n] \\&+ \frac{p_c}{p_n+p_c}E[Y_1(1,0)|D=0,M=0,\tau= c], \\
\stackrel{A7}{=}& \frac{p_n}{p_n+ p_c} E[Y_1(1,0)|\tau=n] + \frac{p_c}{p_n+p_c}E[Y_1(1,0)|\tau= c].
\end{align*}
After some rearrangements and using (\ref{a3}), we obtain
\begin{equation*}
E[Y_1(1,0)|\tau= c]  = \frac{p_n+p_c}{p_c} E[Q_{10}(Y_0)|D=0,M=0] -   \frac{p_n}{p_c} E[Y_1|D=1,M=0].
\end{equation*}
This gives
\begin{equation} \label{p4}
\begin{array}{rl}
E[Y_1(1,0)|\tau=c] =&  \displaystyle \frac{p_{0|0}}{p_{0|0} - p_{0|1}}E[Q_{10}(Y_0)|D=0,M=0] \\ & \displaystyle \quad  - \frac{p_{0|1}}{p_{0|0} - p_{0|1}}E[Y_1|D=1,M=0],
\end{array}
\end{equation}
using $p_n = p_{0|1}$, and $p_c +p_n = p_{0|0}$.

\subsection{Quantile direct effect under $\mathbf{d = 0}$ on compliers}

We show that
\begin{align*}
 F_{Y_{1}(1,0)|\tau=c}(y) &= \frac{p_{0|0}}{p_{0|0} - p_{0|1}}  F_{Q_{10}(Y_{0})|D=0,M=0}(y) - \frac{p_{0|1}}{ p_{0|0} - p_{0|1}c}  F_{Y_1|D=1,M=0}(y) \mbox{ and} \\
F_{Y_{1}(0,0)|\tau=c}(y) &= \frac{p_{0|0}}{p_{0|0} - p_{0|1}} F_{Y_{1}|D=0,M=0}(y) - \frac{p_{0|1} }{p_{0|0} - p_{0|1}}F_{Q_{00}(Y_{0})|D=1,M=0}(y)  ,
\end{align*}
which proves that $\theta_1^{c}(q,0) = F_{Y_{1}(1,0)|c}^{-1}(q)-F_{Y_{1}(0,0)|c}^{-1}(q)$ is identified.

From (\ref{qq3}), we have $F_{Y_{1}(1,0)|D=0,M(0)=0}(y)   = F_{Q_{10}(Y_{0})|D=0,M=0}(y)$.
Applying the law of iterative expectations gives
\begin{align*}
F_{Y_{1}(1,0)|D=0,M(0)=0}(y)   =& \frac{p_n}{p_n + p_c} F_{Y_{1}(1,0)|D=0,M(0)=0,\tau=n}(y)\\ & +\frac{p_c}{p_n + p_c} F_{Y_{1}(1,0)|D=0,M(0)=0,\tau=c}(y),\\
 \stackrel{A7}{=}& \frac{p_n}{p_n + p_c} F_{Y_{1}(1,0)|\tau=n}(y) +\frac{p_c}{p_n + p_c} F_{Y_{1}(1,0)|\tau=c}(y).
\end{align*}
Using (\ref{qq9}) and rearranging the equation gives,
\begin{equation} \label{ff1}
F_{Y_{1}(1,0)|\tau=c}(y) = \frac{p_{0|0}}{p_{0|0} - p_{0|1}}  F_{Q_{10}(Y_{0})|D=0,M=0}(y) - \frac{p_{0|1}}{ p_{0|0} - p_{0|1}}  F_{Y_1|D=1,M=0}(y).
\end{equation}

In analogy to (\ref{a1}), the outcome distribution under $D=0$ and $M=0$ equals:
\begin{equation*}
F_{Y_{1}|D=0,M=0}(y) =\frac{p_n}{p_n + p_c}F_{Y_{1}(0,0)|\tau=n}(y)  + \frac{p_c}{p_n + p_c}F_{Y_{1}(0,0)|\tau=c}(y).
\end{equation*}
Using (\ref{qq10}) and rearranging the equation gives
\begin{equation} \label{ff2}
F_{Y_{1}(0,0)|\tau=c}(y)= \frac{p_{0|0}}{p_{0|0} - p_{0|1}} F_{Y_{1}|D=0,M=0}(y) - \frac{p_{0|1} }{p_{0|0} - p_{0|1}}F_{Q_{00}(Y_{0})|D=1,M=0}(y) .
\end{equation}

\section{Proof of Theorem 4 \label{appE}}

 \subsection{Average direct effect on the
always-takers}

In the following, we show that $\theta_1^a= E[Y_1(1,1)-Y_1(0,1)|\tau=a]= E[Q_{11}(Y_0)-Y_1|D=0,M=1]$. From (\ref{a15}), we obtain the first ingredient $E[Y_1(0,1)|a]=E[Y_1|D=0,M=1]$. Furthermore, from (\ref{q11}) we have $E[Q_{11}(Y_0)|D=0,M=1] = E[Y_1(1,1)|D=0,M(0)=1]$. Under Assumption 7 and 8,
\begin{equation} \label{pp2} \begin{array}{rl} E[Y_1(1,1)|D=0,M(0)=1]&\stackrel{A7}{=}E[Y_1(1,1)|D=0,\tau=a]\\&\stackrel{A8}{=}E[Y_1(1,1)|\tau=a].
\end{array} \end{equation}

 \subsection{Quantile direct effect on the
always-takers}

We prove that
\begin{align*}
\theta_1^a (q) &= F_{Y_1(1,1)|\tau=a}^{-1}(q)- F_{Y_1(0,1)|\tau=a}^{-1}(q), \\
&= F_{Q_{11}(Y_0)|D=0,M=1}^{-1}(q)-F_{Y_1|D=0,M=1}^{-1}(q).
\end{align*}
This requires showing that
\begin{align}
F_{Y_1(1,1)|\tau=a}(y) &=F_{Q_{11}(Y_0)|D=0,M=1}(y) \mbox{ and} \label{qq11}\\
F_{Y_1(0,1)|\tau=a}(y) & = F_{Y_1|D=0,M=1}(y). \label{qq12}
\end{align}
Under Assumptions 7 and 8,
\begin{equation} \label{a15b}
\begin{array}{rl}
F_{Y_t|D=0,M=1} (y) &= E[1\{Y_t\leq y\}|D=0,M=1] \\& \stackrel{A7,A8}{=}E[1\{Y_t(0,1)\leq y\}|\tau=a] \\& = F_{Y_{t}(0,1)|\tau=a}, (y). \end{array}
\end{equation}
which proves (\ref{qq12}). From (\ref{qqq2}), we have
\begin{equation*}
F_{Q_{11}(Y_{0})|D=0,M=1}(y) = F_{Y_{1}(1,1)|D=0,M(0)=1}(y)=E[1\{Y_1(1,1) \leq y\}|D=0,M(0)=1].
\end{equation*}
Under Assumption 7 and 8,
\begin{equation} \label{qqq4} \begin{array}{rl} E[1\{Y_1(1,1) \leq y\}|D=0,M(0)=1]&\stackrel{A7,A8}{=}E[1\{Y_1(1,1) \leq y\}|\tau=a]\\ & = F_{Y_1(1,1)|\tau=a}(y),
\end{array} \end{equation}
which proves (\ref{qq11}).

\subsection{Average direct effect
under $\mathbf{d = 1}$ on compliers}

In the following, we show that
\begin{align*}
\theta_{1}^{c}(1) =& E[Y_1(1,1)-Y_1(0,1)|\tau=c], \\ = &
\frac{p_{1|1}}{ p_{1|1} - p_{1|0}} E[Y_1-Q_{01}(Y_0) |D=1,M=1] \\& -\frac{p_{1|0}}{p_{1|1} - p_{1|0}}E[Q_{11}(Y_0)-Y_1|D=0,M=1].
\end{align*}
Plugging (\ref{pp2}) in (\ref{a1b}), we obtain
\begin{equation*}
\begin{array}{rl}
E[Y_1|D=1,M=1] =& \displaystyle \frac{p_a}{p_a + p_c}E[Q_{11}(Y_0)|D=0,M=1] \\
& \displaystyle \quad + \frac{p_c}{p_a + p_c} E[Y_1(1,1)|\tau=c].
\end{array}
\end{equation*}
This allows identifying
\begin{equation} \label{p2}
\begin{array}{rl}
E[Y_1(1,1)|\tau=c] =& \displaystyle \frac{p_{1|1}}{ p_{1|1} - p_{1|0}} E[Y_1|D=1,M=1] \\
& \displaystyle \quad -\frac{p_{1|0}}{p_{1|1} - p_{1|0}}E[Q_{11}(Y_0)|D=0,M=1].
\end{array}
\end{equation}

From (\ref{q01}) we have $E[Y_1(0,1)|D=1,M=1] = E[Q_{01}(Y_0)|D=1,M=1]$. Applying the law of iterative expectations, gives
\begin{align*}
E[Y_1(0,1)|D=1,M=1] =& \frac{p_a}{p_a+p_c} E[Y_1(0,1)|D=1,M=1, \tau=a] \\&+ \frac{p_c}{p_a+p_c}E[Y_1(0,1)|D=1,M=1,\tau= c], \\
\stackrel{A7}{=}& \frac{p_a}{p_a+ p_c} E[Y_1(0,1)|\tau=a] + \frac{p_c}{p_a+p_c}E[Y_1(0,1)|\tau= c].
\end{align*}
After some rearrangements and using (\ref{a15}), we obtain
\begin{equation*}
E[Y_1(0,1)|\tau= c]  = \frac{p_a+p_c}{p_c} E[Q_{01}(Y_0)|D=1,M=1] -   \frac{p_a}{p_c} E[Y_1|D=0,M=1].
\end{equation*}
This gives
\begin{equation} \label{p3}
\begin{array}{rl}
E[Y_1(0,1)|\tau=c] = & \displaystyle \frac{p_{1|1}}{p_{1|1}-p_{1|0}}E[Q_{01}(Y_0)|D=1,M=1] \\
& \displaystyle \quad - \frac{p_{1|0}}{p_{1|1}-p_{1|0}}E[Y_1|D=0,M=1],
\end{array}
\end{equation}
with $p_a = p_{1|0}$, and $p_c +p_a =p_{1|1}$.

\subsection{Quantile direct effect
under $\mathbf{d = 1}$ on compliers}

We show that
\begin{align*}
 F_{Y_{1}(1,1)|\tau=c}(y) &=  \frac{p_{1|1}}{p_{1|1} - p_{1|0}} F_{Y_{1}|D=1,M=1}(y) - \frac{p_{1|0} }{p_{1|1} - p_{1|0}}F_{Q_{11}(Y_0)|D=0,M=1}(y) \mbox{ and} \\
F_{Y_{1}(0,1)|\tau=c}(y)& = \frac{p_{1|1}}{p_{1|1} - p_{1|0}}  F_{Q_{01}(Y_{0})|D=1,M=1}(y) - \frac{p_{1|0}}{ p_{1|1} - p_{1|0}}  F_{Y_1|D=0,M=1}(y)  ,
\end{align*}
which proves that $\theta_1^{c}(q,1) = F_{Y_{1}(1,1)|c}^{-1}(q)-F_{Y_{1}(0,1)|c}^{-1}(q)$ is identified.

In analogy to (\ref{a1b}), the outcome distribution under $D=0$ and $M=0$ equals:
\begin{equation*}
F_{Y_{1}|D=1,M=1}(y)=\frac{p_a}{p_a + p_c}F_{Y_{1}(1,1)|\tau=a}(y)  + \frac{p_c}{p_a + p_c}F_{Y_{1}(1,1)|\tau=c}(y).
\end{equation*}
Using (\ref{qq11}) and rearranging the equation gives
\begin{equation}
F_{Y_{1}(1,1)|\tau=c}(y)= \frac{p_{1|1}}{p_{1|1} - p_{1|0}} F_{Y_{1}|D=1,M=1}(y) - \frac{p_{1|0} }{p_{1|1} - p_{1|0}}F_{Q_{11}(Y_0)|D=0,M=1}(y) . \label{ff3}
\end{equation}

From (\ref{qq8}), we have $F_{Y_{1}(0,1)|D=1,M(1)=1}(y)  =F_{Q_{01}(Y_{0})|D=1,M=1}(y)$. Applying the law of iterative expectations gives
\begin{align*}
F_{Y_{1}(0,1)|D=1,M(1)=1}(y)   =&\frac{p_a}{p_a + p_c} F_{Y_{1}(0,1)|D=1,M(1)=1,\tau=a}(y)
\\ & +\frac{p_c}{p_a + p_c} F_{Y_{1}(0,1)|D=1,M(1)=1,\tau=c}(y),\\
 \stackrel{A7}{=}& \frac{p_a}{p_a + p_c} F_{Y_{1}(0,1)|\tau=a}(y) +\frac{p_c}{p_a + p_c} F_{Y_{1}(0,1)|\tau=c}(y).
\end{align*}
Using (\ref{qq12}) and rearranging the equation gives,
\begin{equation} \label{ff4}
F_{Y_{1}(0,1)|\tau=c}(y) = \frac{p_{1|1}}{p_{1|1} - p_{1|0}}  F_{Q_{01}(Y_{0})|D=1,M=1}(y) - \frac{p_{1|0}}{ p_{1|1} - p_{1|0}}  F_{Y_1|D=0,M=1}(y).
\end{equation}

\section{Proof of Theorem 5 \label{appF}}

\subsection{Average treatment effect on the compliers}

In (\ref{p2}) and (\ref{p1}), we show that
\begin{align*}
\theta_1^c=&  E[Y_1(1,1) -Y_1(0,0)|\tau=c], \\ =& \frac{p_{1|1}}{ p_{1|1} - p_{1|0}} E[Y_1|D=1,M=1] -\frac{p_{1|0}}{p_{1|1} - p_{1|0}}E[Q_{11}(Y_0)|D=0,M=1] \\
& − \frac{p_{0|0}}{ p_{0|0} - p_{0|1}} E[Y_1|D=0,M=0] +\frac{p_{0|1}}{p_{0|0} - p_{0|1}}E[Q_{00}(Y_0)|D=1,M=0] .
\end{align*}

\subsection{Quantile treatment effect on the compliers}

In (\ref{ff3}) and (\ref{ff2}), we show that $F_{Y_{1}(1,1)|c}(y)$ and $F_{Y_{1}(0,0)|c}(y)$ are identified. Accordingly, $\Delta_1^{c}(q) = F_{Y_{1}(1,1)|c}^{-1}(q)-F_{Y_{1}(0,0)|c}^{-1}(q)$ is identified.

\subsection{Average indirect effect
under $d = 0$ on compliers}

In (\ref{p3}) and (\ref{p1}), we show that
\begin{align*}
\delta_{1}^{c}(0) =& E[Y_1(0,1) -Y_1(0,0)|\tau=c],\\
=& \frac{p_{1|1}}{p_{1|1}-p_{1|0}}E[Q_{01}(Y_0)|D=1,M=1] - \frac{p_{1|0}}{p_{1|1}-p_{1|0}}E[Y_1|D=0,M=1]\\
&- \frac{p_{0|0}}{ p_{0|0} - p_{0|1}} E[Y_1|D=0,M=0] +\frac{p_{0|1}}{p_{0|0} - p_{0|1}}E[Q_{00}(Y_0)|D=1,M=0].
\end{align*}

\subsection{Quantile indirect effect
under $d = 0$ on compliers}

In (\ref{ff4}) and (\ref{ff2}), we show that $F_{Y_{1}(0,1)|c}(y)$ and $F_{Y_{1}(0,0)|c}(y)$ are identified. Accordingly, $\delta_1^{c}(q,0) = F_{Y_{1}(0,1)|c}^{-1}(q)-F_{Y_{1}(0,0)|c}^{-1}(q)$ is identified.

\subsection{Average indirect effect
under $d = 1$ on compliers}

In (\ref{p2}) and (\ref{p4}), we show that
\begin{align*}
\delta_1^{c}(1) =& E[Y_1(1,1) -Y_1(1,0)|\tau= c],\\
=& \frac{p_{1|1}}{ p_{1|1} - p_{1|0}} E[Y_1|D=1,M=1] -\frac{p_{1|0}}{p_{1|1} - p_{1|0}}E[Q_{11}(Y_0)|D=0,M=1]\\
&- \frac{p_{0|0}}{p_{0|0} - p_{0|1}}E[Q_{10}(Y_0)|D=0,M=0] + \frac{p_{0|1}}{p_{0|0} - p_{0|1}}E[Y_1|D=1,M=0].
\end{align*}

\subsection{Quantile indirect effect
under $d = 1$ on compliers}

In (\ref{ff3}) and (\ref{ff1}), we show that $F_{Y_{1}(1,1)|c}(y)$ and $F_{Y_{1}(1,0)|c}(y)$ are identified. Accordingly,  $\delta_1^{c}(q,1) = F_{Y_{1}(1,1)|c}^{-1}(q)-F_{Y_{1}(1,0)|c}^{-1}(q)$ is identified.

\end{appendix}

\end{document}